\documentclass[a4paper]{spie}

\usepackage{graphicx}
\usepackage{mathptmx}
\usepackage[scaled=0.92]{helvet}
\usepackage{courier}
\usepackage[modulo,switch]{lineno}
\usepackage[usenames, dvipsnames]{color}

\newcommand\arcsec{\mbox{$^{\prime\prime}$}}
\newcommand\phn{\phantom{0}}
\newcommand\phb{\phantom{(0)}}

\usepackage{color}
\definecolor{myRed}{rgb}{0.84,0.08,0.52}
\definecolor{black}{rgb}{0,0,0}
\definecolor{blue}{rgb}{0,0,1}
\usepackage{hyperref}
\hypersetup{
    final=true,
    pageanchor=true,
    colorlinks=true,
    breaklinks=true,
    linkcolor=blue,
    citecolor=blue,
    urlcolor=blue,
    pdfpagemode=UseNone,
    pdftitle={GREGOR Fabry-Perot Interferometer - status report and
        prospects},
    pdfauthor={Klaus G. Puschmann, Horst Balthasar, Christian Beck,
        Rohan E. Louis, Emil Popow , Thomas Seelemann, Reiner Volkmer,
        Manfred Woche, and Carsten Denker},
    pdfsubject={Solar Physics},
    pdfkeywords={Sun, spectroscopy, polarimetry, high angular resolution,
        instrumentation, image restoration}}

\title{GREGOR Fabry-P\'erot Interferometer -- status report and prospects}

\author{
    Klaus G. Puschmann\supit{a},
    Horst Balthasar\supit{a},
    Christian Beck\supit{b},
    Rohan E. Louis\supit{a},
    Emil Popow\supit{a},\\
    Thomas Seelemann\supit{c},
    Reiner Volkmer\supit{d},
    Manfred Woche\supit{a}, and
    Carsten Denker\supit{a}
\skiplinehalf
    \supit{a}Leibniz-Institut f\"ur Astrophysik Potsdam,
        An der Sternwarte 16,
        14882 Potsdam,
        Germany\\
    \supit{b}Instituto de Astrof\'isica de Canarias,
        c/ V\'ia L\'actea s/n,
        38205 La Laguna,
        Spain\\
    \supit{c}LaVision,
        Anna-Vandenhoeck-Ring 19,
         37081 G\"ottingen,
         Germany\\
    \supit{d}Kiepenheuer-Institut f\"ur Sonnenphysik,
        Sch\"oneckstra{\ss}e 6,
        79104 Freiburg,
        Germany}

\begin{document}
\maketitle


%
%

\begin{abstract}
The GREGOR Fabry-P\'erot Interferometer (GFPI) is one of three first-light
instruments of the German 1.5-meter GREGOR solar telescope at the Observatorio
del Teide, Tenerife, Spain. The GFPI allows fast narrow-band imaging and
post-factum image restoration. The retrieved physical parameters will be a
fundamental building block for understanding the dynamic Sun and its magnetic
field at spatial scales down to 50~km on the solar surface. The GFPI is a
tunable dual-etalon system in a collimated mounting. It is designed for
spectropolarimetric observations over the wavelength range from 530--860~nm with
a theoretical spectral resolution of ${\cal R}\approx 250,000$. The GFPI is
equipped with a full-Stokes polarimeter. Large-format, high-cadence CCD
detectors with powerful computer hard- and software enable the scanning of
spectral lines in time spans equivalent to the evolution time of solar features.
The field-of-view of $50\arcsec \times 38\arcsec$ covers a significant fraction
of the typical area of active regions. We present the main characteristics of
the GFPI including advanced and automated calibration and observing procedures.
We discuss improvements in the optical design of the instrument and show first
observational results. Finally, we lay out first concrete ideas for the
integration of a second FPI, the Blue Imaging Solar Spectrometer, which will
explore the blue spectral region below 530~nm.
\end{abstract}


%
%

\section{Introduction}\label{SEC01}

Solar physics has made tremendous progress during recent years thanks to
numerical simulations and high-resolution, spectropolarimetric observations with
modern solar telescopes such as the Swedish Solar
Telescope\cite{2003SPIE.4853..341S}, the Solar Optical Telescope on board the
Japanese HINODE satellite,\cite{2007SoPh..243....3K} and the stratospheric
Sunrise telescope.\cite{2010ApJ...723L.127S} Taking the nature of sunspots as an
example, many important new observational results have been found, e.g., details
about the brightness of penumbral filaments, the Evershed flow, the dark-cored
penumbral filaments, the net circular polarization, and the moving magnetic
features in the sunspot moat. Telescopes with apertures of about 1.5~m such as
the GREGOR solar telescope\cite{2007msfa.conf...39V, 2010SPIE.7733E..18V,
2010AN....331..624V, 2012arXiv1202.4289S} or the New Solar
Telescope\cite{2006SPIE.6267E..10D, 2010SPIE.7733E..93C} will help to
discriminate among competing sunspot models and to explain the energy balance of
sunspots. New results on the emergence, evolution, and disappearance of magnetic
flux at smallest scales can also be expected. However, these 1.5-meter
telescopes are just the precursors of the next-generation solar telescopes,
i.e., the Advanced Technology Solar Telescope\cite{2008SPIE.7012E..16W} and the
European Solar Telescope,\cite{2010SPIE.7733E..15C} which will finally be able
to resolve the fundamental scales of the solar photosphere, namely, the photon
mean free path and the pressure scale height.

Fabry-P\'erot interferometers (FPIs) have certainly gained importance in solar
physics during the last decades, because they deliver high spatial and spectral
resolution, and a growing number of such instruments is in operation at various
telescopes. Although most of the instruments have been initially designed only
for spectroscopy, most of them have now been upgraded to provide full-Stokes
polarimetry.\cite{2011A&A...533A..21P, 2011SoPh..268...57M, 2010SPIE.7733E..14R}
The Universit\"ats-Sternwarte G\"ottingen developed an imaging spectrometer for
the German Vacuum Tower Telescope (VTT) in the early 1990s. This instrument used
a universal birefringent filter (UBF) as an order-sorting filter for a
narrow-band FPI mounted in the collimated light beam.\cite{1992A&A...257..817B}
The spectrometer was later equipped with a Stokes-$V$ polarimeter and the UBF
was replaced by a second etalon in 2000.\cite{1995A&A...304L...1V}

A fundamental renewal of the G\"ottingen FPI during the first half of 2005 was
the starting point of the development of a new FPI for the 1.5-meter GREGOR
solar telescope.\cite{2006A&A...451.1151P} New narrow-band etalons and new
large-format, high-cadence CCD detectors were integrated into the instrument,
accompanied by powerful computer hard- and software. From 2006 to 2007, the
optical design for the GREGOR Fabry-P\'erot Interferometer (GFPI) was developed,
the necessary optical elements were purchased, and the opto-mechanical mounts
were manufactured.\cite{2007msfa.conf...45P} An upgrade to full-Stokes
spectropolarimetry followed in 2008.\cite{2008A&A...480..265B,
2009IAUS..259..665B, 2011ASPC..437..351B} In 2009, the Leibniz-Institut f\"ur
Astrophysik Potsdam took over the scientific responsibility for the GFPI, and
the instrument was finally installed at the GREGOR solar
telescope.\cite{2010SPIE.7735E.217D} During the commissioning phase in 2011,
three computer-controlled translation stages (two filter sliders and one mirror
stage) were integrated and the software was prepared for TCP/IP communication
with external devices according to the Device Communication Protocol
(DCP).\cite{2011arXiv1111.5509P} This permits automated observing and
calibration procedures and facilitates easy operations during observing runs.

%
%

\section{GREGOR solar telescope}\label{SEC02}

The 1.5-meter GREGOR telescope is the largest European solar telescope and is
designed for high-precision measurements of dynamic photospheric and
chromospheric structures and their magnetic field. Some key scientific topics of
the GREGOR telescope are: (1) the interaction between convection and magnetic
fields in the photosphere, (2) the dynamics of sunspots and pores and their
temporal evolution, (3) the solar magnetism and its role in solar variability,
and (4) the enigmatic heating mechanism of the chromosphere. The inclusion of a
spectrograph for stellar activity studies and the search for solar twins expand
the scientific usage of the GREGOR telescope to the nighttime
domain.\cite{2007msfa.conf...51S}

The GREGOR telescope replaced the 45-cm Gregory-Coud\'e Telescope, which had
been operated on Tenerife since 1985. The construction of the new telescope was
carried out by a consortium of several German institutes, namely, the
Kiepenheuer-Institut f\"ur Sonnenphysik in Freiburg, the Leibniz-Institut f\"ur
Astrophysik Potsdam (AIP), and the Institut f\"ur Astrophysik G\"ottingen. In
2009, the Max-Planck-Institut f\"ur Sonnensystemforschung in Katlenburg-Lindau
took over the latter contingent. The consortium maintains international
partnerships with the Instituto de Astrof\'isica de Canarias in Spain and the
Astronomical Institute of the Academy of Sciences of the Czech Republic in
Ond\v{r}ejov.

The GREGOR telescope is an alt-azimuthally mounted telescope with an open
structure and an actively cooled light-weighted Zerodur primary mirror. The
completely retractable dome allows wind flushing through the telescope to
facilitate cooling of telescope structure and optics.\cite{2008SPIE.7018E..52B}
The water-cooled field stop at the primary focus provides a field-of-view (FOV)
with a diameter of 150$^{\prime\prime}$. The light is reflected via two
elliptical mirrors, and several flat mirrors along the evacuated coud\'e train
into the optical laboratory. Passing the adaptive optics (AO)
system,\cite{2010ApOpt..49G.155B} the light is finally distributed to the
scientific instruments. The removable GREGOR Polarimetric Unit
(GPU)\cite{2009CEAB...33..317H} is located near the secondary focus and ensures
high-precision polarimetric observations in the visible and near infrared. In
the future, the image rotation introduced by the alt-azimuth mount will be
compensated by a removable image derotator just behind the exit window of the
coud\'e path. A high-order AO system with 196 actuators was recently installed.
The closed-loop bandwidth of the system at 0~dB is 130~Hz. The AO system uses a
Shack-Hartmann wavefront sensor with 156 sub-apertures. A multi-conjugate AO
system will be integrated in the near future.\cite{2010ApOpt..49G.155B}

Three first-light instruments have been commissioned in 2011/12: the GRating
Infrared Spectrograph (GRIS),\cite{2008SPIE.7014E.198C} the Broad-Band Imager
(BBI),\cite{2012arXiv1202.4289S} and the GFPI. A folding mirror deflects the
beam either to BBI or to GRIS/GFPI. The latter two instruments can be used
simultaneously, similar to the multi-instrument setups at the
VTT.\cite{2007msfa.conf...55B} A dichroic beamsplitter directs wavelengths above
650~nm to the spectrograph, whereas all shorter wavelengths are reflected
towards to the GFPI. This beamsplitter can be exchanged with a different
beamsplitter with a cutoff above 800~nm. GFPI and BBI are located in an optical
laboratory on the 5$^\mathrm{th}$ floor of the telescope building. GRIS is
situated one floor below and receives the light by a second folding mirror,
which is placed just behind the slit unit and the slit-jaw imaging system in the
optical laboratory.

%
%

\begin{figure}[t]
\centerline{\includegraphics[width=\textwidth]{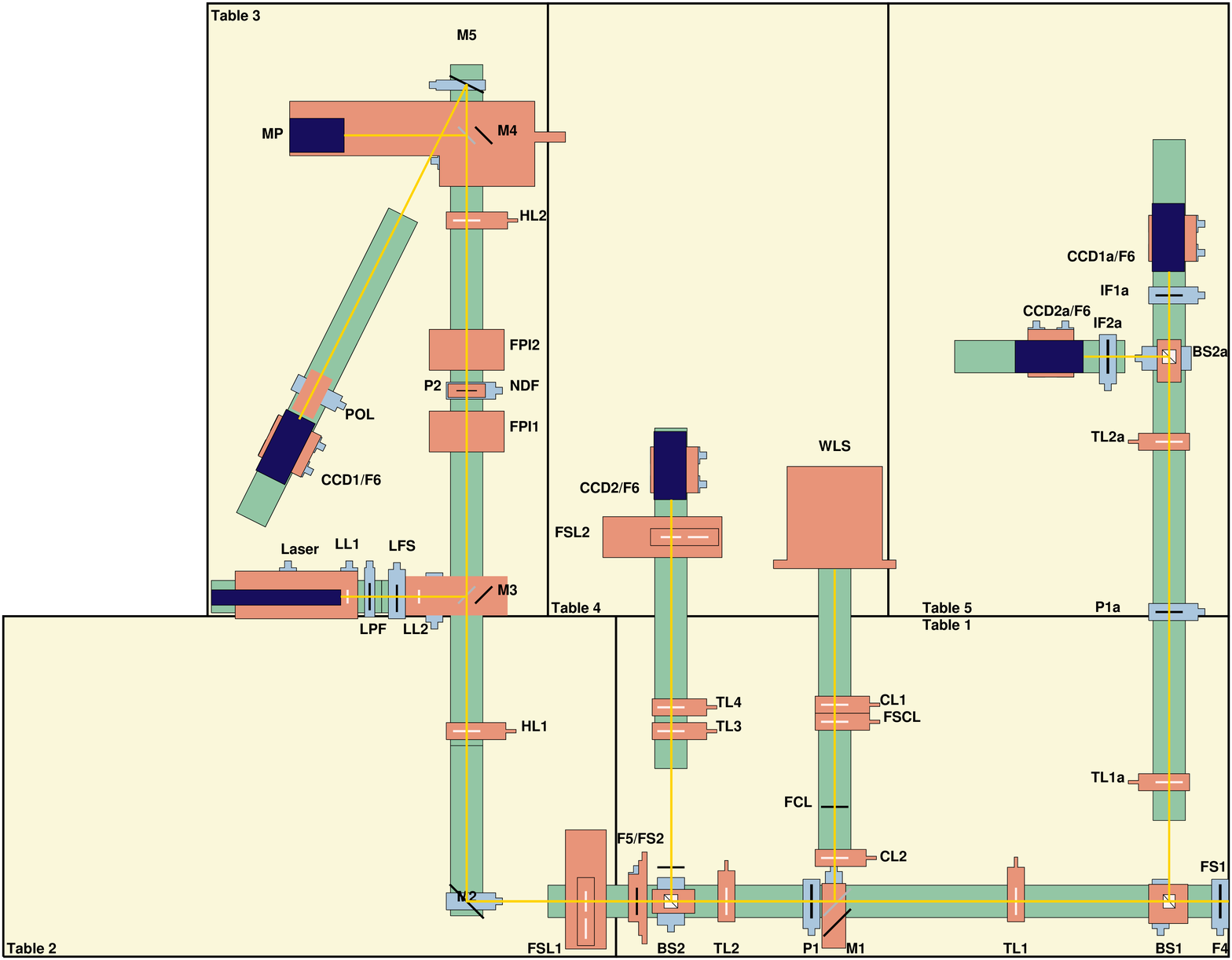}}
\caption{GFPI and blue imaging channel.
    \textsf{CCD1} \& \textsf{CCD2}: Imager QE detectors;
    \textsf{CCD1a} \& \textsf{CCD2a}: pco.4000 detectors;
    \textsf{FPI1} \& \textsf{FPI2}: narrow-band etalons;
    \textsf{NDF}: neutral density filter;
    \textsf{TL1} ($f = 600$~mm, $d = 63$~mm),
        \textsf{TL2} ($f = 250$~mm, $d = 40$~mm),
        \textsf{TL1a} ($f = 500$~mm, $d = 63$~mm),
        \textsf{TL2a} ($f = 500$~mm, $d = 80$~mm),
        \textsf{TL3} ($f = 400$~mm, $d = 63$~mm),
        \textsf{TL4} ($f = 600$~mm, $d = 63$~mm),
        \textsf{HL1} ($f = 1000$~mm, $d = 80$~mm), \&
        \textsf{HL2} ($f = 1500$~mm, $d = 100$~mm): achromatic lenses;
    \textsf{CL1} ($f = 300$~mm, $d = 63$~mm) \&
        \textsf{CL2} ($f = 150$~mm, $d = 40$~mm): plano-convex lenses;
    \textsf{M1}, \textsf{M3}, \& \textsf{M4}: removable folding mirrors;
    \textsf{M2} \& \textsf{M5}: fixed folding mirrors (60~mm $\times$ 85~mm);
    \textsf{F4}, \textsf{F5}, \textsf{F6}, \& \textsf{FCL}: foci;
    \textsf{P1}, \textsf{P1a}, \& \textsf{P2}: pupil images;
    \textsf{BS1}, \textsf{BS2} \& \textsf{BS2a}: beamsplitters (40~mm $\times$
        40~mm);
    \textsf{FS1}, \textsf{FS2} \& \textsf{FSCL}: field stops;
    \textsf{WLS}: white-light source (slide projector);
    \textsf{FSL1} \& \textsf{FSL2}: filter sliders;
    \textsf{IF1a} \& \textsf{IF2a}: interference filters;
    \textsf{POL}: full-Stokes polarimeter;
    \textsf{LL1} \& \textsf{LL2}: laser lenses;
    \textsf{LPF}: laser polarization filter;
    \textsf{LFS}: laser field stop; and
    \textsf{MP}: photomultiplier.}
\label{FIG01}
\end{figure}

\section{GREGOR Fabry-Perot Interferometer}\label{SEC03}

\subsection{Optical design}\label{SEC03_1}

The GFPI is mounted on five optical tables and is protected by an aluminum
housing to prevent the pollution of the optics by dust and to reduce
stray-light. The optical layout of the instrument is shown in Fig.~\ref{FIG01}.
Behind the science focus \textsf{F4}, four achromatic lenses \textsf{TL1},
\textsf{TL2}, \textsf{HL1}, \& \textsf{HL2} create two more foci \textsf{F5} \&
\textsf{F6} and two pupil images \textsf{P1} \& \textsf{P2} in the narrow-band
channel \textsf{NBC}. The etalons \textsf{FPI1} \& \textsf{FPI2} are placed in
the vicinity of the secondary pupil in the collimated light beam. A neutral
density filter \textsf{NDF} between the two etalons with a transmission of 63\%
reduces the inter-etalon reflexes. The beam in the \textsf{NBC} is folded twice
by \textsf{M2} \& \textsf{M5} at a distance of 500~mm and 400~mm from
\textsf{F5} and \textsf{HL2}, respectively, to minimize the instrument envelope.
Two field stops \textsf{FS1} \& \textsf{FS2} reduce stray-light, where the
secondary field stop especially avoids an overlap of the two images created by
the removable dual-beam full-Stokes polarimeter. A beamsplitter cube
\textsf{BS2} near \textsf{F5} directs 5\% of the light to the broad-band channel
\textsf{BBC}. There, the two achromatic lenses \textsf{TL3} \& \textsf{TL4} are
chosen such that the image scale of the detectors \textsf{CCD1} \& \textsf{CCD2}
is exactly the same. A dichroic beamsplitter cube \textsf{BS1} just behind the
science focus \textsf{F4} sends the blue part of the spectrum (below 530~nm) to
the imaging channel. One-to-one imaging with the lenses \textsf{TL1a} \&
\textsf{TL2a} provides the option of recording broad-band images in parallel to
GFPI and GRIS observations with two pco.4000 cameras \textsf{CCD1a} \&
\textsf{CCD2a} behind beamsplitter \textsf{BS2a}. This imaging channel will be
replaced by the Blue Imaging Solar Spectrometer (BLISS) in the future.

Three computer-controlled precision translation stages facilitate automated
observing sequences. The two stages \textsf{FSL1} \& \textsf{FSL2} are used to
switch between two sets of interference filters. The filters restrict the
bandpass for \textsf{BBC} and \textsf{NBC} to a full-width at half-maximum
(FWHM) of 10~nm and 0.3--0.8~nm, respectively. The pre-filters of each channel
can be tilted to optimize the wavelength of the transmission maximum. A third
stage inserts a deflection mirror \textsf{M1} into the light path to take
calibration data with a continuum light source for spectral calibration
purposes. A laser/photo-multiplier channel for finesse adjustment of the etalons
completes the optical setup.

\subsection{Cameras, etalons, and control software}

The GFPI data acquisition system consists of two Imager QE CCD cameras with Sony
ICX285AL detectors, which have a full-well capacity of 18,000~e$^{-}$ and a
read-out noise of 4.5~e$^{-}$. The detectors have a spectral response from
320--900~nm with a maximum quantum efficiency of $\sim$60\% at 550~nm. The chips
have $1376 \times 1040$ pixels with a size of 6.45~$\mu$m $\times$ 6.45~$\mu$m.
The total chip size is 8.8~mm $\times$ 6.7~mm. The image scale at both cameras
is 0.0361\arcsec\ pixel$^{-1}$, which leads to a FOV of $49.7\arcsec \times
37.6\arcsec$ in the spectroscopic mode. The maximal blueshift due to the
collimated mounting of the etalons is about 4.32~pm at 630~nm. The cameras are
triggered by a programmable timing unit sending out analog TTL signals. The
analog-digital conversion is carried out with 12-bit resolution. The data
recorded by the cameras are passed via digital coaxial cables to the GFPI
control computer and are stored on a RAID~0 system.

Two pco.4000 cameras in stock at the observatory can be used for imaging in the
blue channel of the FPI. The pco.4000 camera has a full-well capacity of
60,000~e$^{-}$ and a read-out noise of 11~e$^{-}$. The detector has a spectral
response from 320--900~nm with a quantum efficiency of $\sim$32\% ($\sim$45\%)
at 380 (530)~nm. The pixel size is 9~$\mu$m $\times$ 9~$\mu$m, i.e., $4008
\times 2672$ pixels yield a total chip size of 36~mm $\times$ 24~mm. The image
scale at both cameras is 0.0315\arcsec\ pixel$^{-1}$. To avoid vignetting of the
beam by \textsf{BS1} and \textsf{BS2a} only $2800 \times 2200$ pixels can be
used resulting in a FOV of $88.2\arcsec \times 69.3\arcsec$. Three interference
filters for Ca\,\textsc{ii}\,H $\lambda 396.8$~nm, Fraunhofer G-band $\lambda
430.7$~nm, and blue continuum $\lambda 450.6$~nm with a $\mathrm{FWHM} = 1$~nm
and a transmission better than 60\% are available for observations.

The two GFPI etalons manufactured by IC Optical Systems (ICOS) have a diameter
of $\oslash = 70$~mm, a measured finesse ${\cal F} \sim 46$, spacings $d = 1.1$
and 1.4~mm, and a high-reflectivity coating ($R \sim 95$\%) in the wavelength
range from 530--860~nm. The resulting narrow transmission of the instrument is
on the order of $\mathrm{FWHM} = 1.9$--5.6~pm and leads to a theoretical
spectral resolution of ${\cal R} = 250,000$. All etalons are operated by
three-channel CS100 controllers manufactured by ICOS. The cavity spacings are
digitally controlled by the GFPI control computer via RS-232 communication. A
thermally insulated box protects the pupil and the etalons from stray-light and
air flows inside the instrument.

The communication between internal (cameras, etalons, and filter and mirror
sliders) and peripheral devices (telescope, AO system, AO filter wheel, GPU,
GRIS, etc.) is controlled by the software package DaVis from LaVision in
G\"ottingen, which has been adapted to the needs of the
spectrometer.\cite{2006A&A...451.1151P, 2011arXiv1111.5509P} The modification of
the software for TCP/IP communication with external devices using DCP allows an
easy implementation of automated observing procedures. All observing modes such
as etalon adjustment, line finding, flat-fielding, recording of dark, pinhole,
and target images, continuum scans, and recording of scientific data are now
automated.

\subsection{Polarimetry at GREGOR}

The science verification time in 2012 is mostly devoted to spectroscopy at
GREGOR. Polarimetric observations will follow in 2013. Nevertheless, a
polarization model for the GREGOR telescope has already been
developed,\cite{2011ASPC..437..351B} which is time-dependent because of the
alt-azimuthal mount of the telescope. The GPU\cite{2009CEAB...33..317H} was
developed by and built at AIP and can be inserted at the secondary focus of the
telescope to determine the instrumental polarization, which is important for the
calibration of polarimetric measurements. Since 2008, The GFPI is equipped with
a full-Stokes polarimeter\cite{2008A&A...480..265B} that can be inserted in
front of the detector in the \textsf{NBC}. The polarimeter consists of two
ferro-electric liquid crystal retarders (FLCRs) and a modified Savart-plate. The
first liquid crystal acts as a half-wave plate and the second one as a
quarter-wave plate at a wavelength of 630~nm. The modified Savart-plate consists
of two polarizing beamsplitters and an additional half-wave plate, which
exchanges the ordinary and the extraordinary beam. With this configuration, the
separation of the two beams is optimized and the orientation of the astigmatism
in both beams is the same so that it can be corrected by a cylindrical lens. The
present set of FLCRs yields a good efficiency in the spectral range from
580--660~nm. The integration of an automated calibration procedure for the GFPI
will be an important milestone before starting polarimetric observations.

%
%

\section{GFPI science verification}\label{SEC04}

For a first characterization of the GFPI performance, we took several data sets
in a technical campaign from 15~May to 1~June 2012. The AO system was not
available because of technical problems with the control computer. Thus, our
efforts have been restricted to an optimization of the system and to
observations of test data for an estimate of intensity levels, image quality,
spectral resolution, stray-light, and other performance indicators of the GFPI
and its extended blue imaging  channel. Several problems related to the RS-232
communication with the FPI controllers were resolved so that a stable finesse is
now achieved for several days. In addition, the timing between cameras and
etalons has been optimized to ensure that images are only taken when the etalon
spacing has settled to its nominal value.

\subsection{Imaging spectrometric data}

\begin{table}[t]
\begin{center}
\caption{Summary of the GFPI observations.}
\label{TAB01}
\footnotesize
\medskip
\begin{tabular}{lcccccccccc}
\hline\hline
Channel & $\lambda_0$ [nm] & $\Delta t$ [ms] & $I$ [counts] & \textsf{PHA}
    (\textsf{F4}) & \textsf{PH} (\textsf{F3}) & \textsf{PH} (\textsf{F2}) &
    \textsf{TG} & \textsf{QS} & \textsf{SP} & \textsf{PF}\rule[-2mm]{0mm}{6mm}\\
\hline
NBC ($1 \times 1$ binning) & 543.3 &     40 & \phn 2400 &  x &  x & -- & x & x &  x &  x\rule{0mm}{4mm}\\
BBC ($1 \times 1$ binning) & 543.3 &     40 &        -- &  x &  x & -- & x & x &  x & --\\
NBC ($2 \times 2$ binning) & 543.3 &     30 & \phn 3500 & -- & -- & -- & x & x & -- &  x\\
NBC ($2 \times 2$ binning) & 557.6 &     40 & \phn 2100 & -- &  x & -- & x & x & -- &  x\\
NBC ($2 \times 2$ binning) & 617.3 &     10 & \phn 3200 & -- &  x & -- & x & x & -- &  x\rule[-2mm]{0mm}{4mm}\\
\hline
Ca\,\textsc{ii}\,H & 396.8 &     15 & \phn 9230 &  x &  x &  x & x & x &  x & --\rule{0mm}{4mm}\\
G-band             & 430.7 & \phn 6 &     10350 &  x &  x &  x & x & x &  x & --\\
Blue continuum     & 450.6 & \phn 3 &     11300 &  x &  x &  x & x & x &  x & --\rule[-2mm]{0mm}{4mm}\\
\hline\vspace*{-2mm}
\end{tabular}
\parbox{0.87\textwidth}{Note. --- Central wavelength $\lambda_0$,
    exposure time $\Delta t$, and mean intensity $I$ for all observations
    (\textsf{PHA}: pinhole array, \textsf{PH}: pinhole, \textsf{TG}: target,
    \textsf{QS}: quiet Sun, \textsf{SP}: sunspot, and \textsf{PF}:
    pre-filter).}
\end{center}
\end{table}

\begin{table}[t]
\begin{center}
\caption{NBC pre-filter characteristics.}
\label{TAB02}
\medskip
\footnotesize
\begin{tabular}{lcccccccc}
\hline\hline
Filter\rule[-2mm]{0mm}{6mm} & $\lambda_0$ [nm] & FWHM [nm] & $T$ [\%] & Binning &
   $\Delta T$ [ms] & $I$ [counts] & Frame rate [Hz]\\
\hline
ANDV11436       & 543.4 & 0.4 & 38 & $1 \times 1$ & 60~~~(100)        & 1200~~~(2000)           & \phn 7~~~(5)\rule{0mm}{4mm}\\
                & 543.4 & 0.4 & 38 & $2 \times 2$ & 30~~~\phb\phn\phn & 3500~~~\phb\phn\phn\phn &     11~~~\phb \\
1100 BARR9      & 543.4 & 0.6 & 70 & $1 \times 1$ & 40~~~\phn (60)    & 1700~~~(2700)           & \phn 7~~~(6) \\
ANDV5288        & 557.6 & 0.3 & 40 & $2 \times 2$ & 40~~~\phb\phn\phn & 2100~~~\phb\phn\phn\phn &     11~~~\phb \\
DV5289 AM-32389 & 569.1 & 0.3 & 45 & $1 \times 1$ & 60~~~(100)        & 1200~~~(2000)           & \phn 6~~~(5)\\
ANDV9330        & 617.3 & 0.7 & 80 & $1 \times 1$ & 20~~~\phn (60)    & 1200~~~(3400)           & \phn 9~~~(6)\\
                & 617.3 & 0.7 & 80 & $2 \times 2$ & 10~~~\phb\phn\phn & 3200~~~\phb\phn\phn\phn &     16~~~\phb \rule[-2mm]{0mm}{4mm}\\
\hline\vspace*{-2mm}
\end{tabular}
\parbox{0.87\textwidth}{Note. --- Continuum intensities $I$ at wavelengths
    $\lambda_0$ for filters with peak transmissions $T$ and exposure times
    $\Delta T$.}
\end{center}
\end{table}

Three complete data sets with $2 \times 2$ binning ($\sim$0.0722\arcsec\
pixel$^{-1}$) were taken on 31~May and 1~June, which included images of the
target \textsf{TG} and pinhole \textsf{PH} mounted in the AO filter wheel at
\textsf{F3} (telescope focal plane). The observations covered the spectral lines
at 543.4~nm, 557.6~nm, and 617.3~nm. The Fe\,\textsc{i} $\lambda 543.4$~nm line
had already been scanned on 27~May at full spatial resolution (0.0361\arcsec\
pixel$^{-1}$) including images of a pinhole array in \textsf{F4} (science
focus at the entrance of the GFPI). Simultaneous \textsf{BBC} images were
recorded (Tab.~\ref{TAB01}). Line scans with the GFPI were usually carried out
using a step width of eight digital-analog (DA) steps, whereas the pre-filter
scans were performed with one DA step. One DA step corresponds to 0.26--0.41~pm
in the spectral range from 530--860~nm.

Broad-band data were taken  in the blue imaging channel on 26 and 27~May 2012
just before the GFPI measurements. The observing scheme was the same for all
available pre-filters, i.e., 396.8~nm, 430.7~nm, and 450.6~nm. In addition to a
few observations of the quiet Sun and a sunspot, images of the target and
pinhole in \textsf{F3}, the pinhole-array in \textsf{F4}, and the pinhole
mounted in the GPU at \textsf{F2} are also included. All data in the blue
channel were taken with a combination of two spare lenses with $f = 500$~mm and
1250~mm because a second lens with $f=500$~mm was not available, yet. Thus, the
image scale of 0.0124\arcsec\ pixel$^{-1}$ oversamples the diffraction limit at
these wavelengths by a factor of about three. The proper image scale of
0.0315\arcsec\ pixel$^{-1}$ will be obtained by one-to-one imaging.

\subsection{Intensity estimates for the narrow-band channel}

The wavelength $\lambda_0$, FWHM, and transmission $T$ of different \textsf{NBC}
filters are summarized in Tab.~\ref{TAB02} together with the selected binning,
the counts at continuum wavelengths, and frame rates at a given exposure time
$\Delta T$. The results reveal that at full spatial resolution very long
exposure times of up to 100~ms are necessary to achieve at least 2000 counts in
the continuum of most of the measured spectral lines for most of the filters
with $T \sim 40$\% and a $\mathrm{FWHM} \sim 0.3$--0.4 nm. As a consequence, one
can reach only very low frame rates. A $2 \times 2$-pixel binning speeds up the
observations and reduces the exposure times. The situation changes when choosing
filters with higher transmission. The 617.3~nm filter with $T \sim 80$\% and a
$\mathrm{FWHM} \sim 0.74$~nm yields reasonable frame rates and exposure times
even without binning.

\begin{figure}[t]
\includegraphics[width=0.49\textwidth]{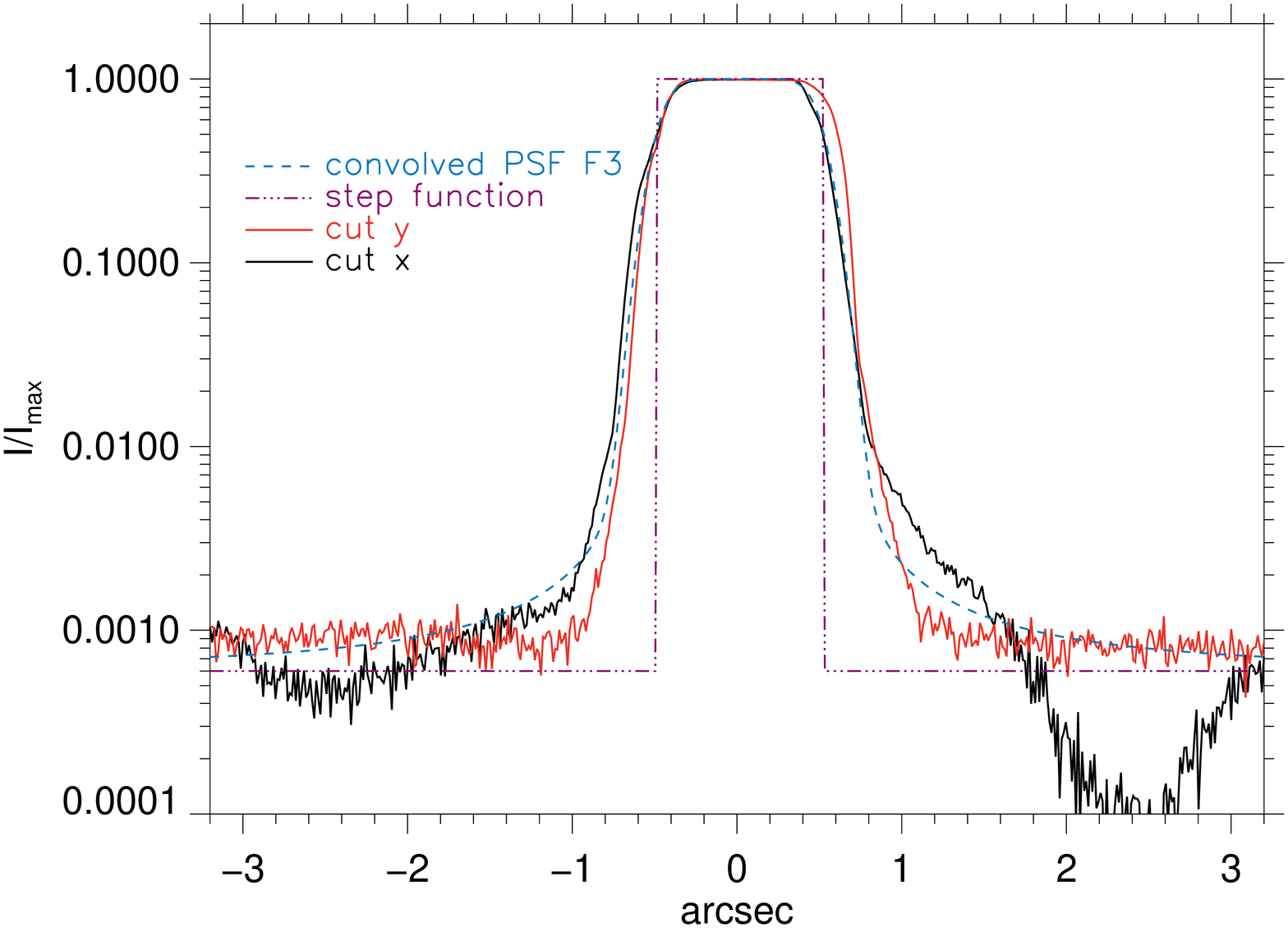}
\hfill
\includegraphics[width=0.49\textwidth]{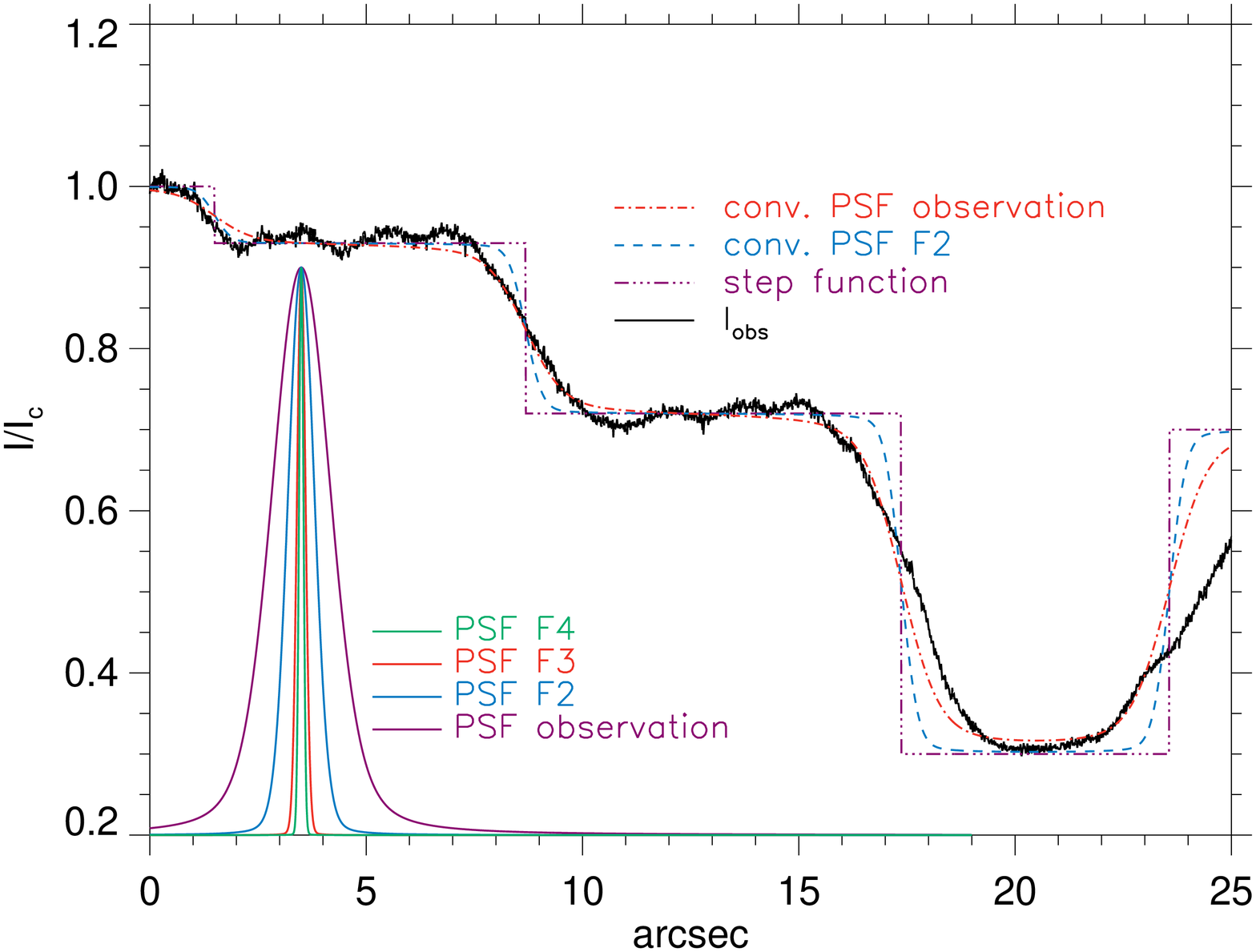}
\caption{Derivation of the PSF estimate at 430~nm for the pinhole in \textsf{F3}
    (\textit{left}) and for the complete optical train from a sunspot
    observation (\textit{right}), where the PSF estimates of the four different
    focal planes are displayed in the lower left corner.}
\label{FIG02}
\end{figure}

\subsection{Estimates of the spatial point spread function}

Knowledge of the instrumental point spread function (PSF) provides an estimate
of both the spatial resolution and the \textit{spatial} stray-light level
expected for an instrument or telescope.\cite{2011A&A...535A.129B,
2012A&A...537A..80L} Using a reference such as a pinhole or a blocking edge in
the focal plane, the PSF of all optics downstream can be derived. The combined
PSF of the telescope, the post-focus instruments, and the time-variable seeing
can also be derived from the observations of a sunspot with its steep spatial
intensity gradients.

The PSF of the optical train at the GREGOR telescope relevant for the GFPI and
its complementary imaging channels was calculated based on reduced images of
pinholes located in the focal planes \textsf{F4}, \textsf{F3}, \textsf{F2}, and
on sunspot images (see Tab.~\ref{TAB01}). All available calibration images,
e.g., target or pinhole images, were averaged for each wavelength step or image
burst. However, only a single image was selected for solar observations because
the seeing varies during the image sequences. The pinhole images were normalized
to the maximum intensity $I_{\rm max}$ inside the FOV near the center of the
pinhole, whereas the sunspot data were normalized to unity in a quiet Sun region
outside the spot. All data were taken without real-time correction of the AO
system and consequently correspond to the static performance of the optical
system.

\subsection{Derivation of the point spread function estimates}

We took cuts along the $x$- and $y$-axes of the CCD across the center of the
pinhole in each pinhole observation to obtain an estimate of the PSF. We defined
a step function that is assumed to represent the physical extent of the true
pinhole. The borders of the step function were set to intersect the observed
intensity along the cuts at about the 50\% level (left panel of
Fig.~\ref{FIG02}). We constructed a convolution kernel from a combination of a
Gaussian (variance $\sigma$) and a Lorentzian function (parameter $a$) and
convolved the step function with the kernel. A modification of the parameters
($\sigma, a$) within a specific range yielded finally the kernel that best
matched the convolved step function and the observed intensity along the
horizontal and vertical cuts. The cuts in $x$ and $y$ differed far away from the
pinhole because of the read-out direction of the CCD camera. Thus, we always
tried to find a good match to both the $x$- and $y$-cuts close to the pinhole.
However, we concentrated only on matching the cut perpendicular to the CCD
read-out direction away from the pinhole. The same method was applied to all
pinhole observations (\textsf{F4}, \textsf{F3}, and \textsf{F2}) to derive a PSF
estimate for each focal plane and wavelength. The sunspot observations were
modeled by a similar step function at the two transitions from quiet Sun to
penumbra and penumbra to umbra (right panel of Fig.~\ref{FIG02}).

The resulting PSF estimates of the four focal planes for the imaging data at
430.7~nm are displayed in the lower left corner of the right-hand panel of
Fig.~\ref{FIG02}. The width of the PSF estimates is similar for \textsf{F4} and
\textsf{F3}, where only static optical components and negligible seeing effects
contribute to the PSF and a diffraction-limited performance can be reached, but
increases significantly when passing to the focal plane \textsf{F2} that
experiences telescope seeing and seeing fluctuations along the coud\'e train.
The PSF derived from the sunspot observations includes all seeing effects and
roughly doubles its width relative to the PSF at \textsf{F2}.

\begin{figure}[t]
\begin{minipage}[c]{0.44\textwidth}
\includegraphics[width=\textwidth]{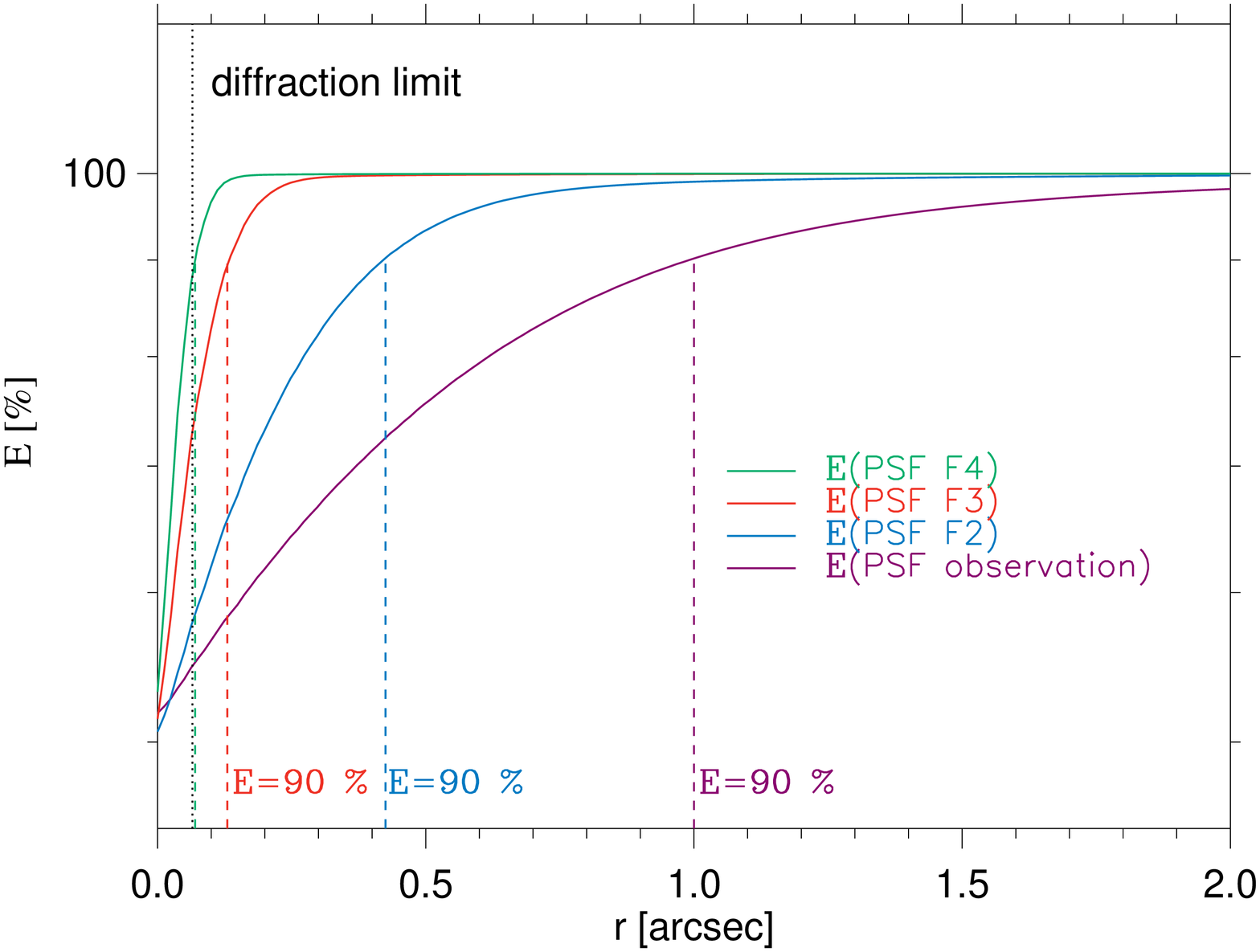}
\end{minipage}
\begin{minipage}[c]{0.56\textwidth}
\begin{scriptsize}
\begin{tabular}{lccccccc}
\hline\hline
Channel               & \multicolumn{4}{c}{Imaging $\lambda 430.7$~nm}        & \multicolumn{3}{c}{BBC $\lambda 543.4$~nm}\rule[-1.5mm]{0mm}{4.5mm} \\
\hline
Sampling              & \multicolumn{4}{c}{0.0124\arcsec\ pixel$^{-1}$}       & \multicolumn{3}{c}{0.034\arcsec\ pixel$^{-1}$}\rule{0mm}{3mm} \\
                      & \multicolumn{4}{c}{(0.0126\arcsec\ pixel$^{-1}$)}     & \multicolumn{3}{c}{(0.036\arcsec\ pixel$^{-1}$)} \\
Diff.\ limit          & \multicolumn{4}{c}{0.072\arcsec}                      & \multicolumn{3}{c}{0.091\arcsec}\rule[-1.5mm]{0mm}{3mm} \\
\hline
Focal plane           & \textsf{F4} & \textsf{F3} & \textsf{F2} & Solar obs.   & \textsf{F4} & \textsf{F3} & Solar obs.\rule[-1.5mm]{0mm}{4.5mm} \\
\hline
$r$ ($E = 90\%$)      & 0.07\arcsec & 0.13\arcsec & 0.43\arcsec & 1.00\arcsec & 0.11\arcsec & 0.17\arcsec & 2.4\arcsec\rule{0mm}{3mm} \\
$E$ ($r_\mathrm{DL}$) &        90\% &        73\% &        58\% &        55\% &        85\% &        74\% & 52\%\rule[-1.5mm]{0mm}{3mm} \\
\hline
\end{tabular}\\
\medskip

\begin{tabular}{lccccc}
\hline\hline
Channel            & \multicolumn{3}{c}{NBC $\lambda 543.4$~nm} & NBC $\lambda 557.6$~nm & NBC $\lambda 617.3$~nm\rule[-1.5mm]{0mm}{4.5mm} \\
\hline
Sampling           & \multicolumn{3}{c}{0.034\arcsec\ pixel$^{-1}$}   & 0.069\arcsec\ pixel$^{-1}$ & 0.067\arcsec\ pixel$^{-1}$\rule{0mm}{3mm} \\
                   & \multicolumn{3}{c}{(0.036\arcsec\ pixel$^{-1}$)} & (0.072\arcsec\ pixel$^{-1}$) & (0.072\arcsec\ pixel$^{-1}$) \\
Diff.\ limit       & \multicolumn{3}{c}{0.091\arcsec}                 & 0.094\arcsec & 0.103\arcsec \rule[-1.5mm]{0mm}{3mm} \\
\hline
Focal plane        & \textsf{F4} & \textsf{F3} & Solar obs.   & \textsf{F3} & \textsf{F3} \rule[-1.5mm]{0mm}{4.5mm} \\
\hline
$r(E = 90\%)$      & 0.12\arcsec & 0.20\arcsec & 2.65\arcsec & 0.19\arcsec & 0.19\arcsec \rule{0mm}{3mm} \\
$E(r_\mathrm{DL})$ &        85\% &        71\% &        51\% &        75\% &        76\% \rule[-1.5mm]{0mm}{3mm} \\
\hline
\end{tabular}
\end{scriptsize}
\end{minipage}\medskip
\caption{Energy enclosed within the radius $r$ for the PSF estimates at
    430.7~nm of the four different focal planes (\textit{left}). The
    \textit{dashed vertical lines} denote the radius where 90\% of the energy
    is enclosed. The \textit{dotted vertical line} denotes the diffraction
    limit at 430.7~nm. Similar observations were carried out at different
    wavelength (\textit{right}). Theoretical values are given in parentheses.}
\label{FIG03}
\end{figure}

By applying the approach described above to all available data, PSF estimates
for all four focal planes were obtained in case of the imaging data at 430.7~nm,
and for some of the focal planes in case of the GFPI \textsf{BBC} and
\textsf{NBC} data (see table in Fig.~\ref{FIG03}). The optical performance and
the expected stray-light level can best be quantified from the total energy
enclosed in a given radius, i.e., a radial integration of the PSF whose volume
yields the amount of light spread up to a given radius. Figure~\ref{FIG03} shows
the enclosed energy $E$ for the PSF estimates at 430.7~nm. From the curves, we
derived two values, namely, the energy enclosed at the radius $E(r_\mathrm{DL})$
of the diffraction limit and the radius $r(E = 90\%)$ at which 90\% of the
energy is enclosed. The former value provides an estimate of the generic
stray-light to be expected by subtracting it from 100\%. The latter value can be
used as generic estimate of the spatial resolution. The table in
Fig.~\ref{FIG03} lists the two values for all analyzed wavelengths and channels,
i.e., the blue imaging channel, and the GFPI \textsf{BBC} and \textsf{NBC} at
specific wavelengths.

A comparison between the value of the diffraction limit and the radius where
90\% of the energy is enclosed shows that the optics behind \textsf{F4} performs
close to diffraction limit. The optics downstream of \textsf{F3} performs
slightly worse with a total enclosed energy of about 70--75\% of the
diffraction-limited case. This implies a generic spatial stray-light level -- if
one defines stray-light as all light scattered to outside a distance of one
times the diffraction limit -- of about 25\% created by the optics downstream of
\textsf{F3}. The corresponding value at \textsf{F4} is about 10--15\%. The
values of both the stray-light level and the radius that encloses 90\% of the
total energy experience a profound jump when passing to \textsf{F2} and beyond.
All these data were taken at mediocre seeing conditions and without AO
correction. The latter will be necessary for a final characterization of the
optical performance of the complete optical train including the telescope.

\subsection{Spectral resolution, spectral stray-light, and blue-shift}

\begin{figure}[t]
\includegraphics[width=0.49\textwidth]{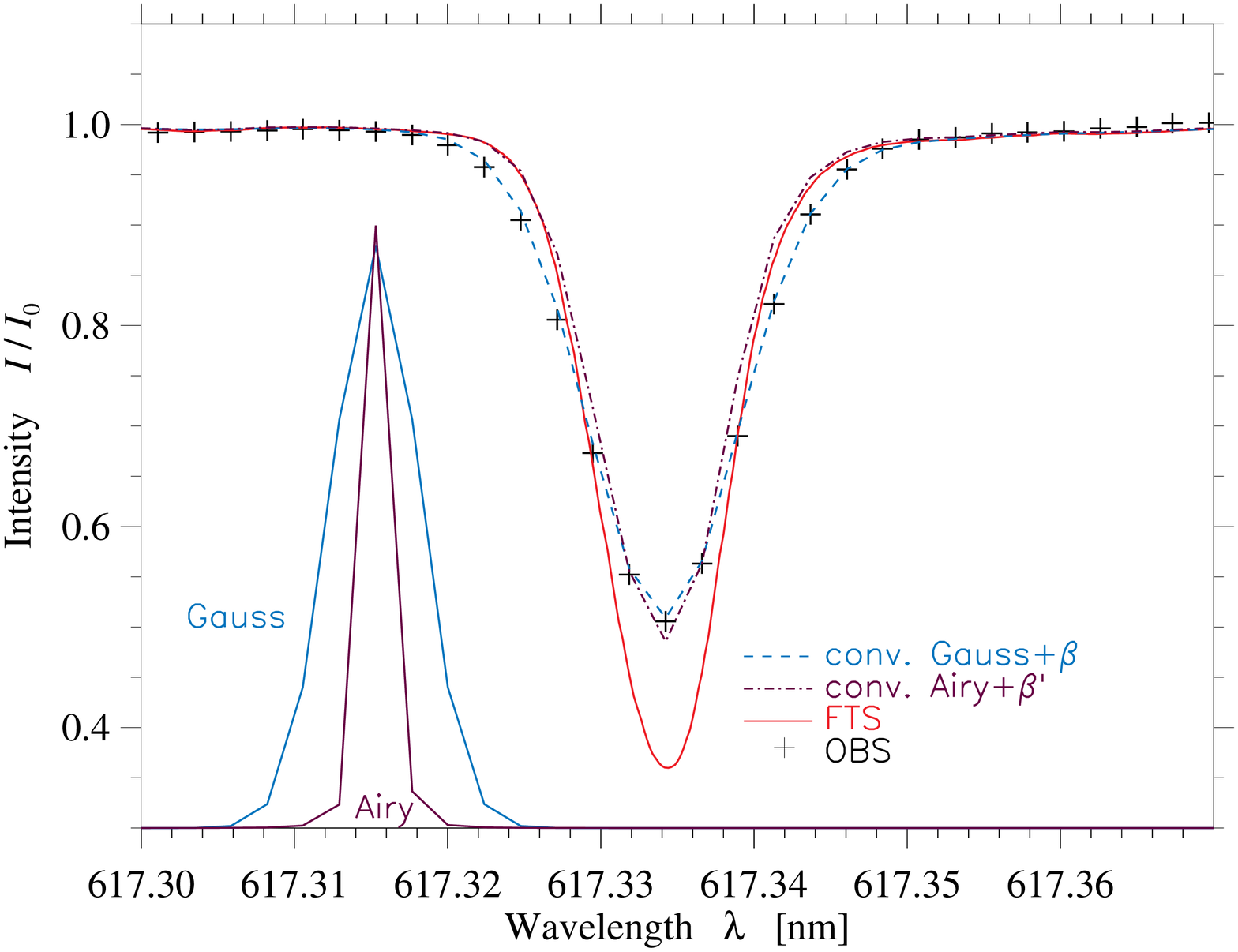}
\hfill
\includegraphics[width=0.49\textwidth]{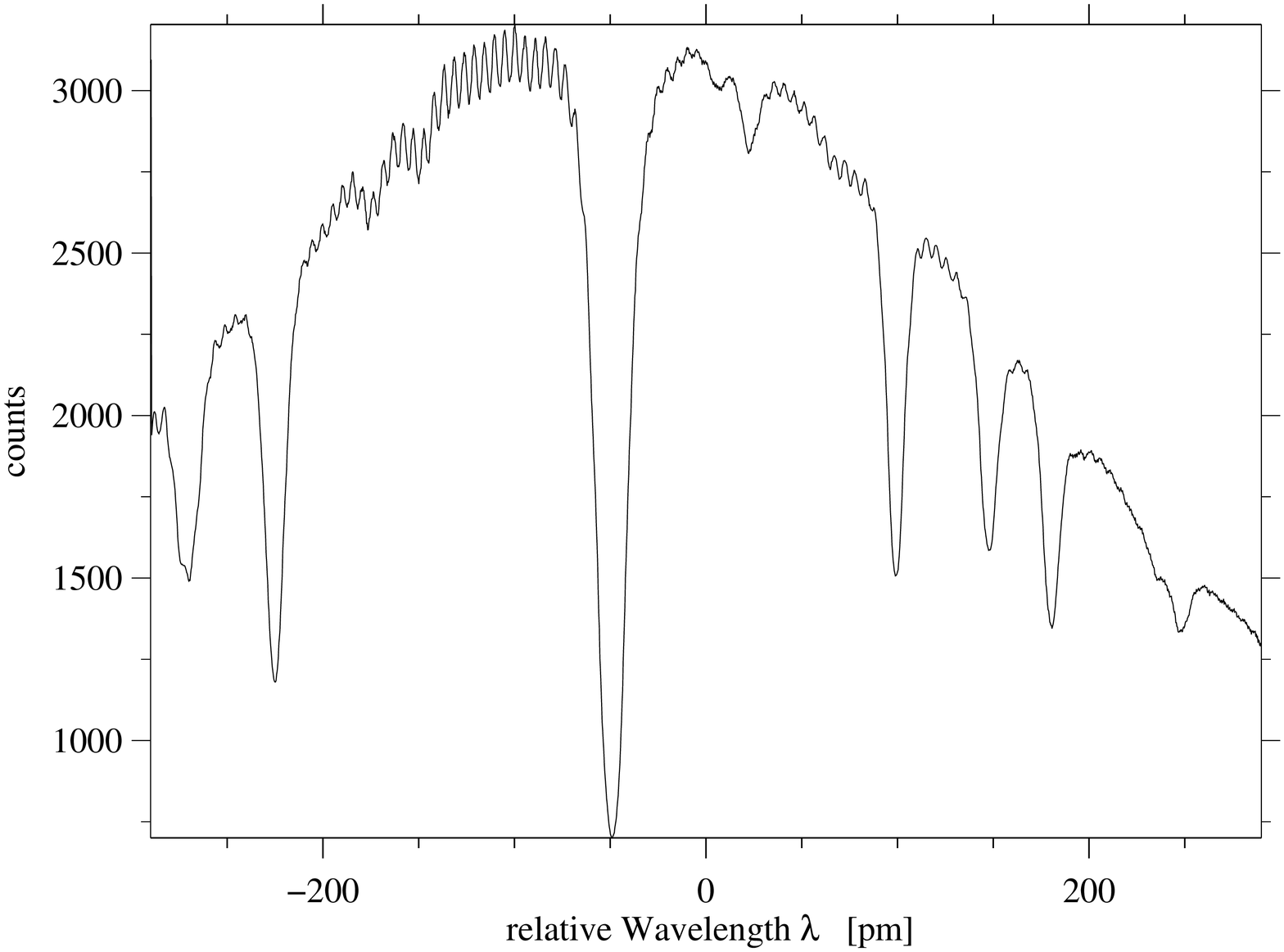}
\caption{Close-up of the Fe\,\textsc{i} $\lambda 617.3$~nm spectral line
    at reduced spectral sampling (\textit{left}): observed spectrum
    (\textit{black pluses}), FTS atlas spectrum (\textit{red solid line}), FTS
    convolved with the theoretical Airy function (\textit{orange dash-dotted
    line}), and FTS convolved with the best-fit Gaussian kernel (\textit{blue
    dashed line}). An appropriate stray-light offset ($\beta^\prime, \beta$) was
    added after the convolution. The Airy function and the Gaussian kernel in
    the left corner are displayed in arbitrary units. Scan of the pre-filter
    curve around the $\lambda 543.4$~nm line at maximal spectral sampling
    (\textit{right}).}
\label{FIG04}
\end{figure}

The spectral resolution and the \textit{spectral} stray-light inside the
\textsf{NBC} was estimated by a convolution of Fourier Transform Spectrograph
(FTS) atlas spectra with a Gaussian of width $\sigma$ and a subsequent addition
of a constant wavelength-independent stray-light offset
$\beta$.\cite{2004A&A...423.1109A, 2007A&A...475.1067C} This component of
stray-light corresponds to light scattered onto the CCD detector without being
spectrally resolved. Therefore, it changes the line depth of observed spectral
lines. The convolved FTS spectra in each wavelength range were compared with two
sets of spatially averaged observational profiles that either covered the full
pre-filter transmission curve or only the line inside the same range that is
usually recorded in science observations (543.4~nm, 557.6~nm, 617.3~nm). The
left panel of Fig.~\ref{FIG04} shows the average observed spectrum at
Fe\,\textsc{i} $\lambda 617.33$~nm, the original FTS spectrum, and the FTS
spectrum after the convolution with the best-fit Gaussian kernel and the
addition of the stray-light offset. The method has some ambiguity between
$\sigma$ and $\beta$, which can be modified in opposite directions over some
range near the best-fit values without significantly degrading the reproduction
of the observed spectra. Therefore, the values listed in Tab.~\ref{TAB03} have
an error of about $\pm 0.5$~pm in $\sigma$ and $\pm 5$\% in $\beta$. The
stray-light level $\beta$ inside the GFPI is found to be below 10\% and the
spectral resolution is ${\cal R} \sim 100,000$, quite below the theoretically
expected value of ${\cal R} \sim 250,000$. The dispersion values derived from
the observed spectra are listed in Tab.~\ref{TAB03} and correspond to eight DA
steps. They match the theoretically expected values given in parentheses.

\begin{table}[t]
\begin{center}
\caption{Spectral characteristics of the GFPI data.}
\label{TAB03}
\footnotesize
\medskip
\begin{tabular}{cccccccc}
\hline\hline
$\lambda$ & Dispersion (8~DA) & Dispersion (1~DA) & $\sigma$ & $\beta$ &
$\lambda/\sigma$ & $\lambda/\Delta\lambda$ & Max.\ blue-shift\rule{0mm}{4mm}\\
\rule[-2mm]{0mm}{4mm}[nm]  &        [pm] &    [pm] & [pm] & [\%] &     -- &         -- & [pm]\\
\hline
543.4 & 2.08\phn\phn (2.09) & (0.261) & 2.62 &  6.6 & 207189 & \phn 88166 & 3.91\phn\phn (3.72)\rule{0mm}{4mm}\\
557.6 & 2.15\phn\phn (2.15) & (0.268) & 2.07 &  6.5 & 269453 &     114661 & 3.96\phn\phn (3.82)\\
617.3 & 2.36\phn\phn (2.36) & (0.296) & 2.81 &  7.8 & 219695 & \phn 93487 & 4.49\phn\phn (4.23)\rule[-2mm]{0mm}{4mm}\\
\phn\phn 630.25\cite{2011A&A...533A..21P} & 2.31\phn\phn (2.41) & & 1.65 & 14 &  381818 & 162476 & \phn\phn\phn\phn\phn\phn (4.32)\rule[-2mm]{0mm}{4mm}\\
\hline\vspace*{-2mm}
\end{tabular}
\parbox{0.81\textwidth}{Note. --- Theoretical values are indicated in parentheses.}
\end{center}
\end{table}

For comparison, the FTS spectrum was convolved with the theoretical Airy
function that describes the transmission through the double-etalon system using
the measured finesse of 46. In this case, the line depth after the convolution
could be adjusted by a stray-light offset $\beta^\prime$ of about 20\%, but the
observed line width in the wings of the line is not matched (Fig.~\ref{FIG04}).
The performance of the GFPI in terms of the spectral resolution depends on the
exact alignment of the plate parallelism of the etalons. For any detailed
modeling of the spectral transmission it seems recommended to use a comparison
of observed and atlas spectra instead of relying on the theoretically expected
spectral PSF.

The measured finesse was obtained with a laser light source that produces a beam
of about 15~mm diameter passing through the etalons. Therefore, the parallelism
of the etalon plates was only optimized in the central part. In case of the
observed spectra, the actual pupil diameter of \textsf{P2} in the GFPI is about
60~mm and samples nearly the full surface of the etalons, in contrary to the VTT
setup, where the pupil in the vicinity of the etalons still had a diameter of
only about 40~mm. In the latter case, a spectral resolution still of ${\cal R}
\sim 160,000$ was obtained applying the same method (see Tab.~\ref{TAB03}).

A degradation of the finesse between the central and peripheral regions of about
1.3 was found for a single etalon for radii of 15~mm and
50~mm.\cite{2005PASP..117.1435D} The spectral resolution in the central area of
the etalons of the GFPI at GREGOR would therefore be  ${\cal R} \sim 170,000$,
which is clearly above the thermal line width ($\lambda/\sigma_{th} \approx
134,000$). Note, however, that in this estimate similar effects of the second
etalon are not yet considered. Thus, the effective spectral resolution of the
central area might be even close to the value expected from theory. A widening
of the laser beam, and consequently the adjustment of the finesse over a broader
area fraction, might significantly improve the results.

The maximal blueshift induced in the spectra because of placing the FPIs inside
a collimated beam was determined from a set of flat-field data. The numbers are
close to the theoretically expected values (last column of Tab.~\ref{TAB03}).

\begin{table}[t]
\begin{center}
\caption{Selected spectral lines in the range
    380--500~nm.}\label{TAB04}
\scriptsize
\smallskip
\begin{tabular}{lccccl}
\hline\hline
Ion &
Wavelength $\lambda$ &
Excitation potential $\chi$ &
Land{\'e}-factor $g$ &
Equivalent width &
Comment \rule[-1.5mm]{0mm}{4.5mm}\\
\hline
Fe\,\textsc{i}     & 384.998~nm & \phn 1.01~eV & 0.00 & \phn\phn    60.8~pm & blends \rule{0mm}{3mm} \\
\textbf{CN-band}   & 388.300~nm &              &      &                     & \\
He                 & 388.865~nm &     19.73~eV &      &                     & only prominences\\
H$\zeta$           & 388.905~nm &     10.20~eV &      & \phn       234.6~pm & only prominences\\
Fe\,\textsc{i}     & 392.922~nm & \phn 3.25~eV & 2.50 & \phn\phn\phn 3.7~pm & weak line \\
Ca\,\textsc{ii}\,K & 393.368~nm & \phn 0.00~eV &      &           2025.3~pm & \\
Ca\,\textsc{ii}\,H & 396.849~nm & \phn 0.00~eV &      &           1546.7~pm & \\
H$\epsilon$        & 397.008~nm &     10.20~eV &      &                     & line wing Ca\,\textsc{ii}\,H \\
Fe\,\textsc{i}     & 404.583~nm & \phn 1.48~eV &      & \phn       117.5~pm & blends \\
Fe\,\textsc{i}     & 406.539~nm & \phn 3.43~eV & 0.00 & \phn\phn\phn 6.4~pm & \\
Mn\,\textsc{i}     & 407.028~nm & \phn 2.19~eV & 3.33 & \phn\phn\phn 6.6~pm & high priority\\
Sr\,\textsc{ii}    & 407.772~nm & \phn 0.00~eV &      & \phn\phn    42.8~pm & resonance line\\
Fe\,\textsc{i}     & 408.088~nm & \phn 3.29~eV & 3.00 & \phn\phn\phn 6.1~pm & high priority\\
H$\delta$          & 410.175~nm &     10.20~eV &      & \phn       313.3~pm & \\
Fe\,\textsc{ii}    & 430.318~nm & \phn 2.70~eV & 1.47 & \phn\phn     10.3pm & in G-band \\
\textbf{G-band}    & 430.500~nm &              &      &                     & \\
H$\gamma$          & 434.048~nm &     10.20~eV &      & \phn       285.5~pm & \\
Fe\,\textsc{i}     & 440.476~nm & \phn 1.56~eV &      & \phn\phn    89.8~pm & low priority\\
Ba\,\textsc{ii}    & 455.404~nm & \phn 0.00~eV &      & \phn\phn    15.9~pm & resonance line\\
Mg\,\textsc{i}     & 457.110~nm & \phn 0.00~eV &      & \phn\phn\phn 9.2~pm & resonance line\\
Fe\,\textsc{i}     & 461.321~nm & \phn 3.29~eV & 0.00 & \phn\phn\phn 6.6~pm & near $g_\mathrm{eff} = 2.5$ line\\
Cr\,\textsc{i}     & 461.367~nm & \phn 0.96eV  & 2.50 & \phn\phn\phn 6.2~pm & near $g_\mathrm{eff} = 0$ line\\
Ti\,\textsc{i}     & 464.519~nm & \phn 1.73~eV & 2.50 & \phn\phn\phn 1.6~pm & enhanced in spots\\
Fe\,\textsc{i}     & 470.495~nm & \phn 3.69~eV & 2.50 & \phn\phn\phn 5.8~pm & high priority\\
H$\beta$           & 486.134~nm &     10.20~eV &      & \phn       368.0~pm & \\
Fe\,\textsc{i}     & 470.178~nm & \phn 3.93~eV & 1.50 & \phn\phn\phn 4.4~pm & near $g_\mathrm{eff} = 0$ line\\
Ni\,\textsc{i}     & 491.203~nm & \phn 3.77~eV & 0.00 & \phn\phn\phn 4.7~pm & \rule[-1.5mm]{0mm}{3mm} \\
\hline
\end{tabular}
\end{center}
\end{table}

The scans of the pre-filter transmission curve at maximal spectral sampling
showed another spectral feature in the GFPI data whose origin remains unclear up
to now. A beat with stable period and varying amplitude is superimposed on the
spectral lines (right panel of Fig.~\ref{FIG04} for 543.4~nm). This beat is
observed in all spectra, independent of wavelength and pre-filter, and has been
present since the integration of the second narrow-band etalon in 2007. Owing to
the aforementioned modifications, the beat is now absolutely stable with time in
position and amplitude for each filter. Forward or backward scanning of the
spectrum yields identical results. Thus, a removal of the beat is
straightforward using the white-light source as a reference.

%
%

\section{BLue Imaging Solar Spectrometer}

The spatial resolution of a telescope scales inversely proportional with the
observed wavelength. Therefore, observations at short wavelengths (below 530~nm)
offer the opportunity to obtain data with higher spatial resolution. This
advantage is partly diminished by the seeing degradations and the smaller number
of photons at shorter wavelengths. There are relatively few ground-based
instruments for spectral observations in the blue spectral
region.\cite{1993ApJ...414..345L, 2002A&A...389. 1020R, 2005A&A...437.1159B,
2008A&A...479..213B} All of these observations were carried out with
slit-spectrographs that permit only limited improvements by post-factum
restoration techniques.\cite{2011A&A...535A.129B} A Fabry-P\'erot-based imaging
spectrometer provides data suitable for sophisticated post-factum image
restoration techniques to yield high spatial and spectral resolution data. This
is the motivation for BLISS, which will supplant the blue imaging channel of the
GFPI in the near future.

\subsection{Interesting spectral lines in the blue part of the visible spectrum}

\begin{figure}[t]
\centerline{\includegraphics[width=0.8\textwidth]{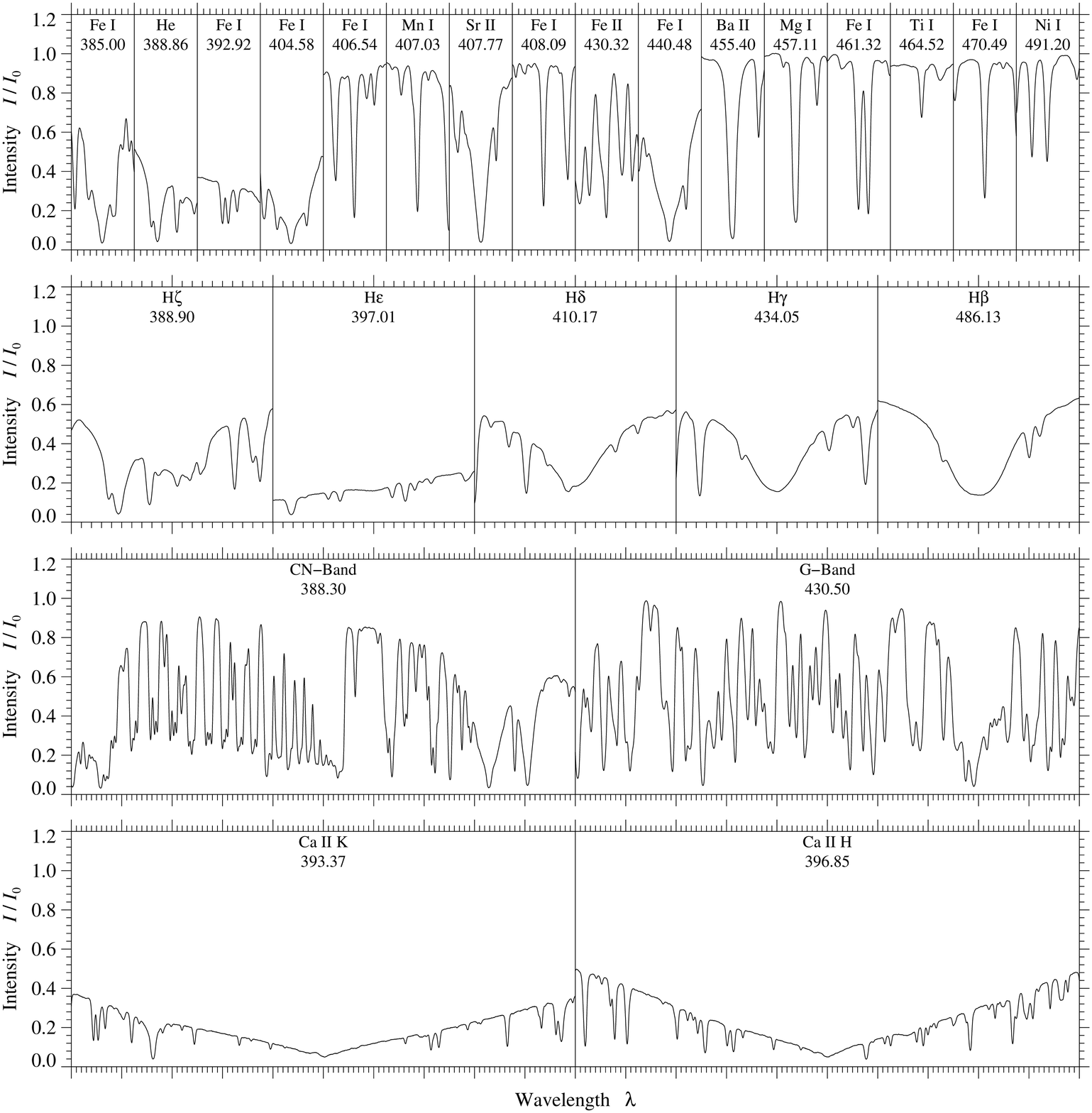}}
\caption{Selected spectral lines in the range 380--500~nm (see
    Tab.~\ref{TAB04}). The name and central wavelength of all spectral lines are
    given at the top of each panel. Major (minor) tick marks are separated by
    0.1~nm (10~pm). Note that the spectral coverage of the panels changes from
    row to row, i.e., 0.1~nm, 0.2~nm, 1.0~nm, and 1.0~nm (\textit{from top to
    bottom}).}
\label{FIG05}
\end{figure}

In the wavelength range covered by BLISS, there are several spectral lines and
two molecular bands of high scientific interest (Tab.~\ref{TAB04} and
Fig.~\ref{FIG05}). All Balmer-lines except H$\alpha$ are at shorter wavelengths
than 530~nm, and the H and K lines of ionized calcium are the strongest lines in
the visible part of the solar spectrum probing the chromosphere. Several
photospheric resonance lines such as Mg\,\textsc{i} $\lambda 457$~nm,
Sr\,\textsc{ii} $\lambda 407$~nm, and Ba\,\textsc{ii} $\lambda 455$~nm are found
in blue part of the visible spectrum. Here, one also finds several magnetically
insensitive lines ($g_\mathrm{eff} = 0$)\cite{1970SoPh...12...66S} and on the
other hand lines exhibiting a Zeeman-triplet with splitting factors of
$g_\mathrm{eff} = 2.5$ and higher.\cite{1973SoPh...28....9H} A special case is
the pair Fe\,\textsc{i} $\lambda 461.32$~nm ($g_\mathrm{eff} = 0$) and
Cr\,\textsc{i} $\lambda 461.37$~nm ($g_\mathrm{eff} = 2.5$) that can be recorded
at the same time. The molecular bands of CH $\lambda 430$~nm (G-band) and CN
$\lambda 388$~nm are well suited to investigate very small hot features in the
solar photosphere.

\subsection{Optical design}

\begin{figure}[t]
\centerline{\includegraphics[width=\textwidth]{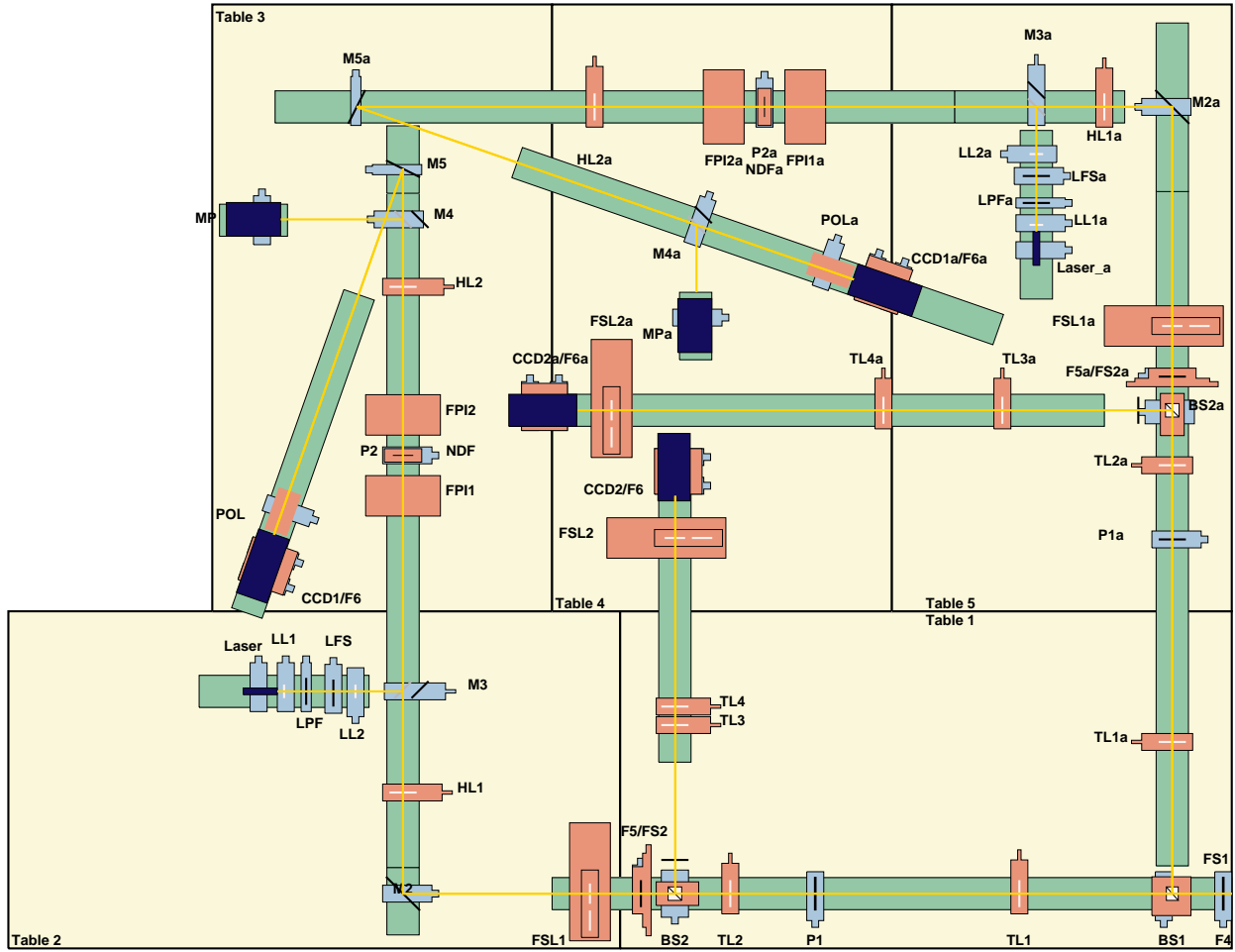}}
\caption{Geometrical design for the integration of BLISS into the GFPI system.
    The design is based on the study presented in Fig.~\ref{FIG07}. The optical
    elements of BLISS are labeled with an extra ``a''. Apart from \textsf{TL2a},
    \textsf{HL2a}, and \textsf{TL4a}, they are identical to those of the GFPI.}
\label{FIG06}
\end{figure}

\begin{figure}[t]
\begin{center}
\includegraphics[width=0.98\textwidth]{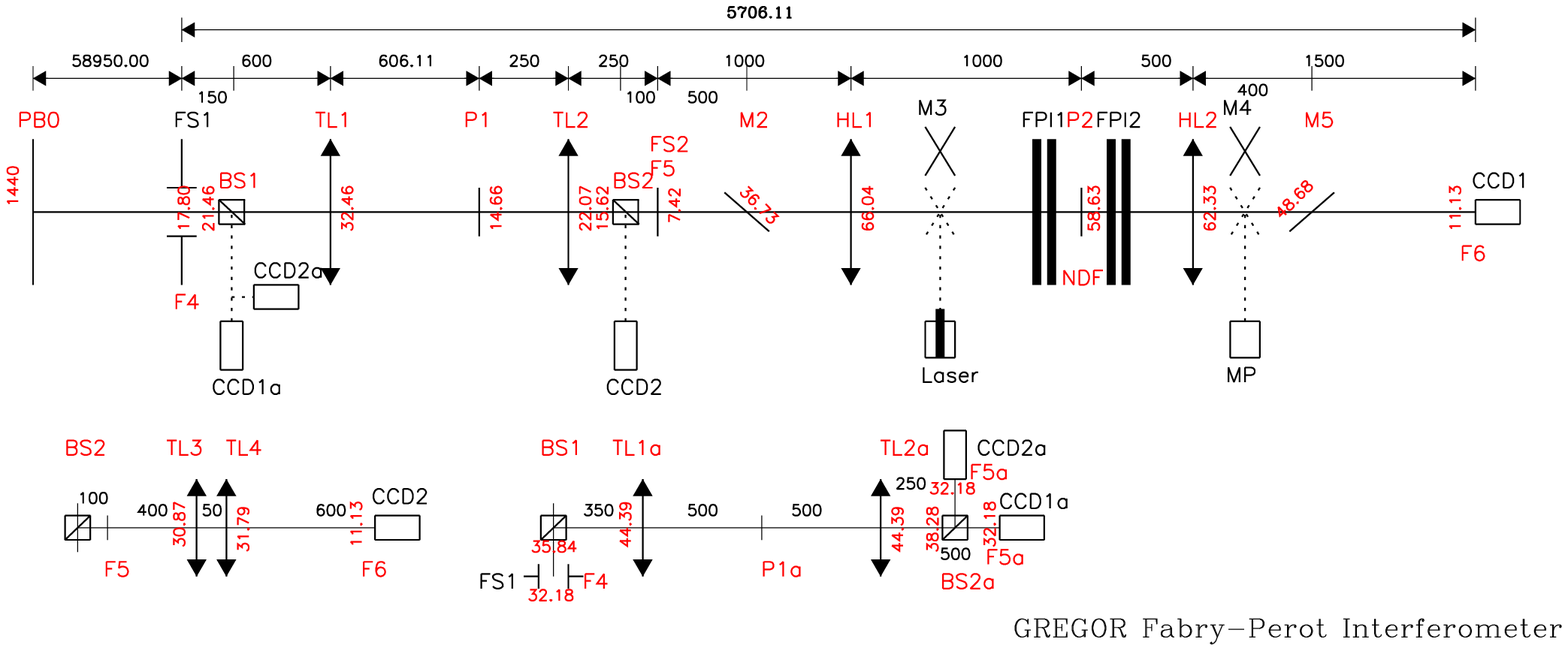}
\includegraphics[width=0.98\textwidth]{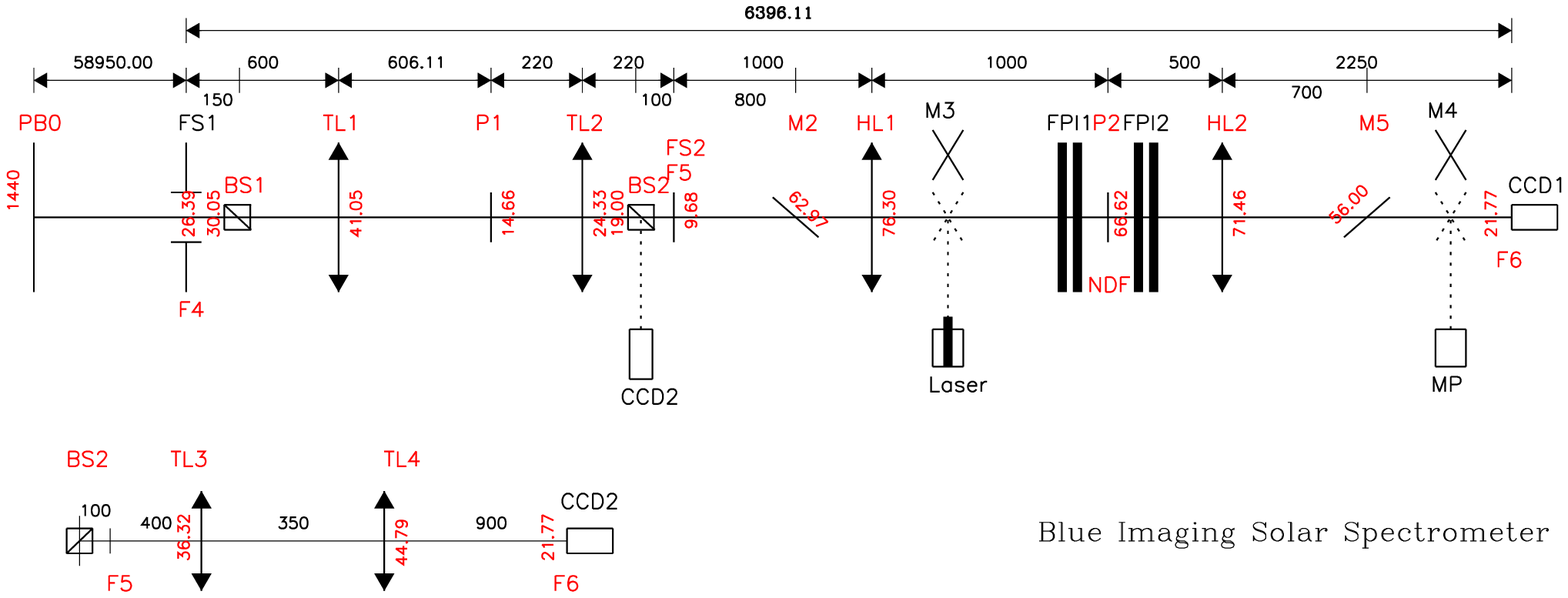}
\end{center}\vspace*{-12mm}
\caption{Schematical optical design (\textit{top}) of the current GFPI
    \textsf{NBC}, \textsf{BBC}, and blue imaging channel. In the future, the
    BLISS \textsf{NBC} and \textsf{BBC} (\textit{bottom}) will replace the GFPI
    blue imaging channel. The diameter of the beam at all optical surfaces, the
    focal length of the lenses, and the positions of other optical elements are
    given in millimeter. The achromatic lenses \textsf{TL2}, \textsf{HL2}, and
    \textsf{TL4} of BLISS have a focal length of 220~mm, 2250~mm, and 900~mm,
    respectively. The mirrors \textsf{M2} and \textsf{M5} of the GFPI will be
    moved for the integration of BLISS by 200~mm and 50~mm, respectively.}
\label{FIG07}
\end{figure}

\begin{figure}[t]
\centerline{\includegraphics[width=\textwidth]{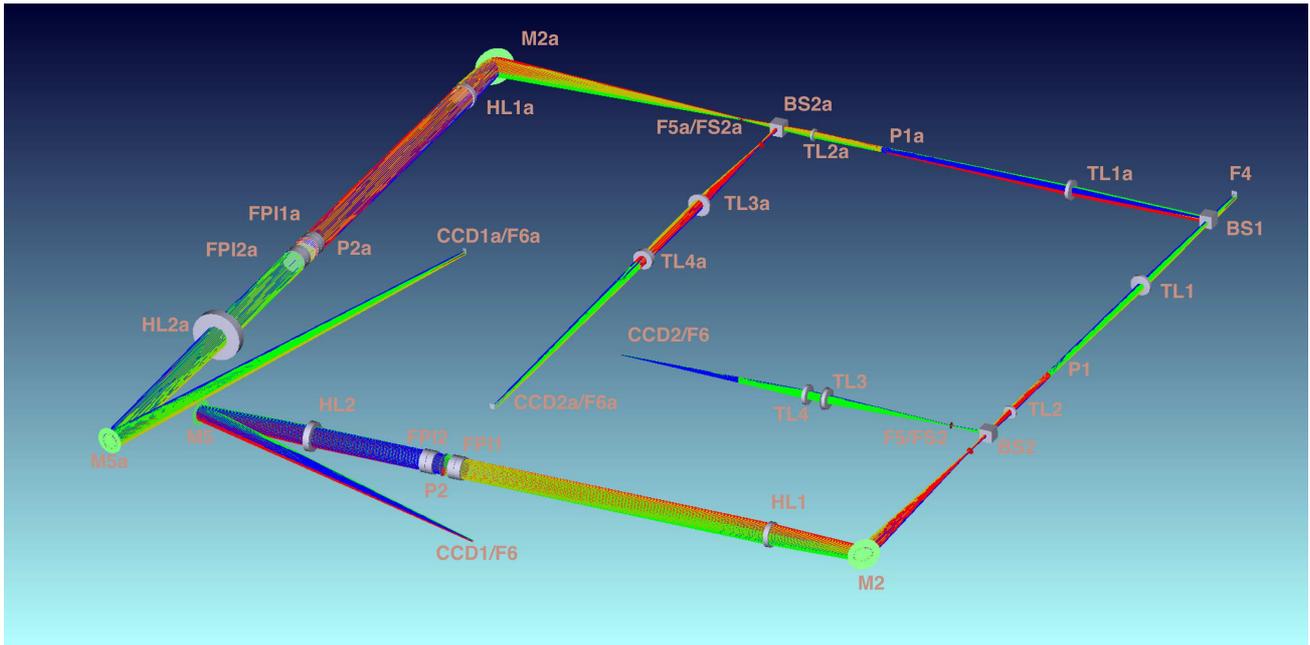}}
\medskip
\caption{Total arrangement of GFPI and BLISS in a ZEMAX multi-configuration
    file in shaded modeling for the respective central wave length of each
    instrument and their maximal field dimension. The design confirms the
    calculations in geometrical optics presented in Figs.~\ref{FIG06} and
    \ref{FIG07}.}
\label{FIG08}
\end{figure}

The geometrical design of BLISS is depicted in Fig.~\ref{FIG06}, which shows its
integration into the GFPI by replacing the blue imaging channel. Most of the
optical components and their labels are identical to those in Fig.~\ref{FIG01}
so that any information can be directly taken from there. The calculations for
this initial optical design revealed that a pupil \textsf{P2} of about 70~mm
diameter will be sufficient for an acceptable maximal blue-shift over the entire
wavelength range at a given image scale and FOV. Thus, the design of BLISS
itself is almost identical to the design of the GFPI, apart from a re-design of
the camera lenses \textsf{HL2} and \textsf{TL4} in the \textsf{NBC} and
\textsf{BBC} to obtain an adequate image scale. The schematical optical designs
of both instruments are compared in Fig.~\ref{FIG07}. The diameter of the light
beam at the different foci, pupils, and on the relevant optical surfaces has
been calculated by means of geometrical optics and confirmed by ZEMAX
ray-tracing as for the GFPI.\cite{2007msfa.conf...45P} The ZEMAX ray-tracing is
presented in Fig.~\ref{FIG08}.

If one considers the actual values of the usable diameter of the main mirror and
the focal length of the GREGOR telescope, the size of the pupil \textsf{P2}
inside the GFPI (59~mm) is smaller than the value assumed in the original
calculation in 2007 (63~mm). However, a modification of the parabolic collimator
and camera mirrors of the AO system will bring the pupil diameter again closer
to the original value. Nevertheless, by changing the focal length of
\textsf{TL2}, the reduction of the pupil diameter is currently compensated in
the design of BLISS (see Fig.~\ref{FIG07}).

A new type of cameras has been envisaged for BLISS, namely, two sCMOS-cameras
from PCO (pco.edge). These cameras have $2560 \times 2160$ pixels with a pixel
size of 6.5~$\mu$m $\times$ 6.5~$\mu$m similar to the GFPI Imager QE cameras. A
change of the focal lengths of \textsf{HL2} from $f = 1500$~mm to $f = 2250$~mm
and of \textsf{TL4} from $f = 600$~mm to $f = 900$~mm for BLISS (see
Fig.~\ref{FIG07}) yields an image scale of 0.0276\arcsec\ pixel$^{-1}$ and a FOV
of $70.6\arcsec \times 59.6\arcsec$ on both cameras. With this configuration,
the maximal blue-shift across the FOV amounts to 4.44~pm and 6.19~pm at 380~nm
and 530~nm, respectively.

In addition, we checked the possibility of interchanging the sCMOS and Imager QE
cameras between the GFPI and BLISS. The almost identical pixel sizes of both
cameras facilitates this task. For the integration of the sCMOS cameras into the
GFPI, a circular FOV with a diameter of $d = 78.8\arcsec$ and an image scale of
0.0364\arcsec\ pixel$^{-1}$ for spectroscopy would result in a maximal blueshift
of 6.91~pm at 630~nm. However, the FOV for polarimetry is limited by the
full-Stokes polarimeter and would remain at its previous size of $24.9\arcsec
\times 37.6\arcsec$. The frame rate at full resolution would increase from 7 to
40~Hz, which perfectly suits this observing mode.

The integration of the Imager QE cameras into BLISS would result in an image
scale of 0.0274\arcsec\ pixel$^{-1}$ and a FOV of $37.6\arcsec \times
28.5\arcsec$ with a maximum blueshift of 1.15~pm and 1.61~pm at 380~nm and
530~nm, respectively. The lower frame rates would match the longer exposure
times at shorter wavelength. A super-achromatic optical setup for BLISS -- as
also foreseen for the GFPI -- would be preferable but otherwise all lenses
except \textsf{TL4} of BLISS could be purchased as off-the-shelf achromats.

The distribution of the two instruments on the five optical tables in the GREGOR
observing room is shown in Fig.~\ref{FIG06}. All optical elements of BLISS have
been labeled with an additional ``a'' to distinguish them from those of the
GFPI. Behind the common dichroic beamsplitter cube \textsf{BS1}, each instrument
has its own \textsf{NBC} and \textsf{BBC}. A displacement of the folding mirrors
\textsf{M2} and \textsf{M5} of the GFPI to a distance of 700~mm and 350~mm from
\textsf{F5} and \textsf{HL2}, respectively, yields some free space on optical
table 3 that can be used for the \textsf{NBC} of BLISS. The ZEMAX ray-tracing
revealed changes in the optical path when considering the etalons plates in the
design. Thus, the GFPI \textsf{NBC} has to be shortened further by reducing the
distance between \textsf{P2} and \textsf{HL2} from 500~mm to 350~mm. This detail
is not considered in Figs.~\ref{FIG06} and \ref{FIG07}.

The beam in the \textsf{NBC} of BLISS is also folded twice by \textsf{M2a} and
\textsf{M5a} at a distance of 800~mm and 700~mm from \textsf{F5a} and
\textsf{HL2a}, respectively. In the \textsf{BBC} of BLISS, \textsf{TL3a} and
\textsf{TL4a} are separated by 350~mm, in contrary to the 50~mm between
\textsf{TL3} and \textsf{TL4} in case of the GFPI. Two computer-controlled
filter sliders \textsf{FSL1a} and \textsf{FSL2a} will again switch between two
sets of interference filters, which restrict the bandpass for the BLISS
\textsf{BBC} and \textsf{NBC}. BLISS is mainly designed for spectroscopy because
of the expected low photon numbers in the blue spectral region. Nevertheless, a
full-Stokes polarimeter can easily be integrated into the system. As in case of
the GFPI, a laser/photo-multiplier channel for finesse adjustment of the etalons
will be implemented. The white-light channel of the GFPI will be removed and be
replaced by an external white-light source common to all instruments at GREGOR.

\subsection{Camera system}\label{SEC05_3}

Modern sCMOS cameras are capable of delivering high frame rates using
large-format sensors with low readout noise, which makes them ideally suited for
an application in solar physics. The pco.edge is a potential candidate for
BLISS. The sensor of this camera  has $2560 \times 2160$ pixels with a size of
6.5~$\mu$m $\times$ 6.5~$\mu$m and a full well capacity of 30,000~e$^{-}$. The
camera has a 16-bit digitization and would be operated in a global shutter mode.
In this mode, the camera has a maximum frame rate of 40~Hz. The cameras are
running in a ``fast scan mode'' (286~MHz) because of speed limitations of the
camera link interface. The 16-bit signal is internally converted to 12-bit in
the fast scan mode. Inside the interface it is decompressed again to 16-bit. The
signal losses due to the compression are roughly a factor of ten smaller than
the shot noise of the camera signal. The readout noise is in the order of
2.3~e$^{-}$. The dark current in the global shutter mode consists of a part
related to the exposure time, i.e., 2--6~e$^{-}$ pixel$^{-1}$ s$^{-1}$ and a
part related to the sensor readout time, which is constant for a given pixel
clock, i.e., 0.6 e$^{-}$ pixel$^{-1}$ in the fast scan mode. Peltier cooling of
the sensor ensures an operating temperature of $+5^{\circ}$~C. The camera has a
quantum efficiency of $\sim$30\% and $\sim$54\% at 380~nm and 530~nm,
respectively, similar to the Imager QE cameras currently used in the GFPI
($\sim$36\% and $\sim$60\%).

To handle the extremely large data bandwidth of about 300~MB~s$^{-1}$ at frame
rates of 40~Hz, each camera will be controlled by an individual PC, in which the
data will be stored on local RAID~0 systems. The integration of the cameras in
the control software is straightforward because the DaVis software will be
operational on the new system with minor changes only. The cameras are currently
being tested at AIP including, e.g., an analysis of image quality, power
spectra, and noise characteristics.

However, the feasibility of using the pco.edge for BLISS has still to be
demonstrated. A detailed photon statistic in the wavelength range
380--530~nm will help with the final decision, if these cameras will be
employed in BLISS or the GFPI. For the blue wavelength range, rather long
exposure times can be expected, making the use of high-speed cameras somewhat
doubtful. Moreover, the exposure time of the cameras currently has an upper
limit of 100 ms due to a relatively high dark current in the global shutter
mode. On the other hand, high frame rates would be extremely beneficial when
operating the GFPI in the vector polarimetric mode, because at present this
instrument is limited to just 5--7~Hz at full resolution. The disadvantage would
be just a partial usage of the chip. The almost identical pixel size facilitates
interchanging the cameras between the two instruments.

\begin{figure}[t]
\includegraphics[width=0.49\textwidth]{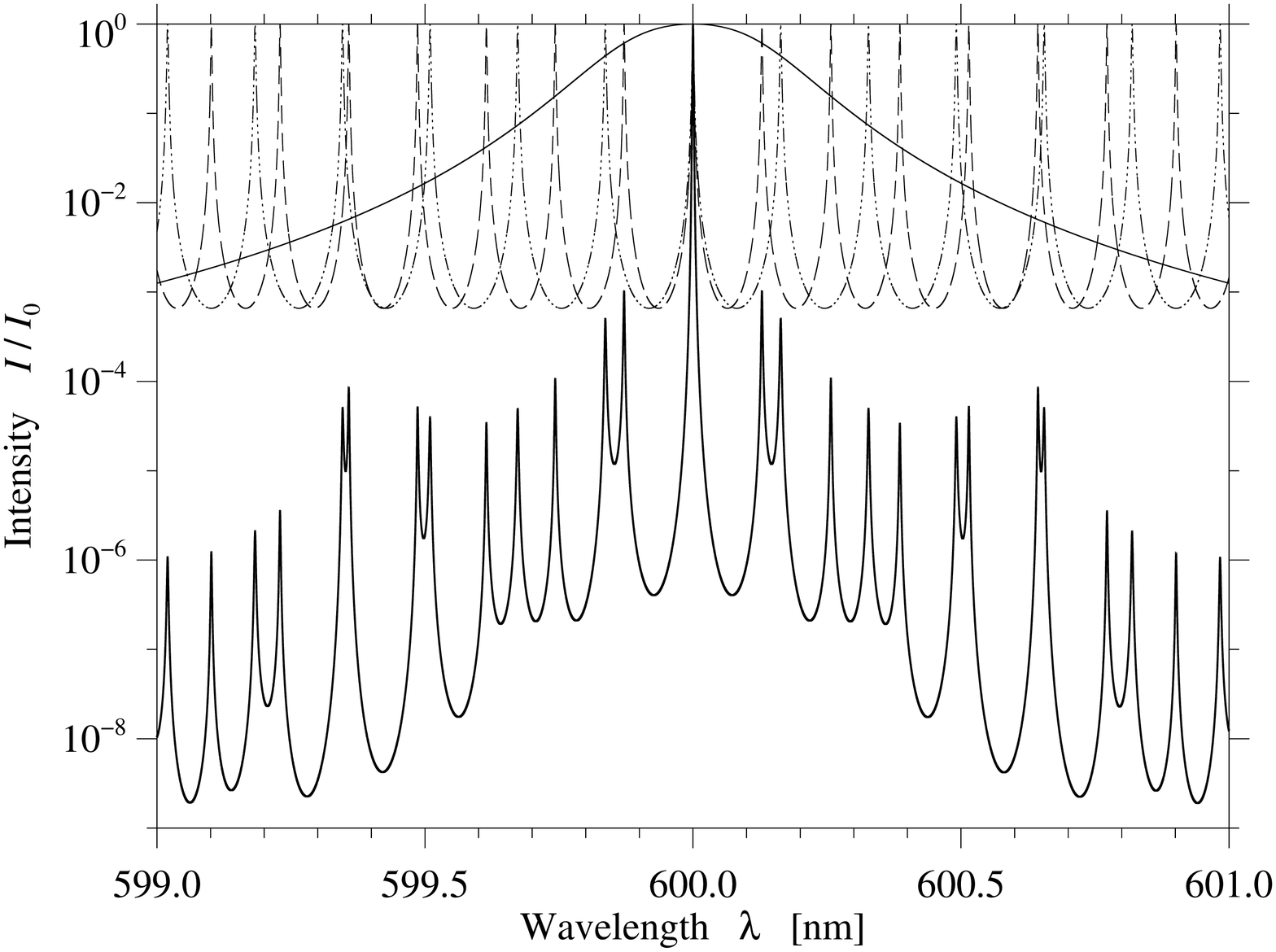}
\hfill
\includegraphics[width=0.49\textwidth]{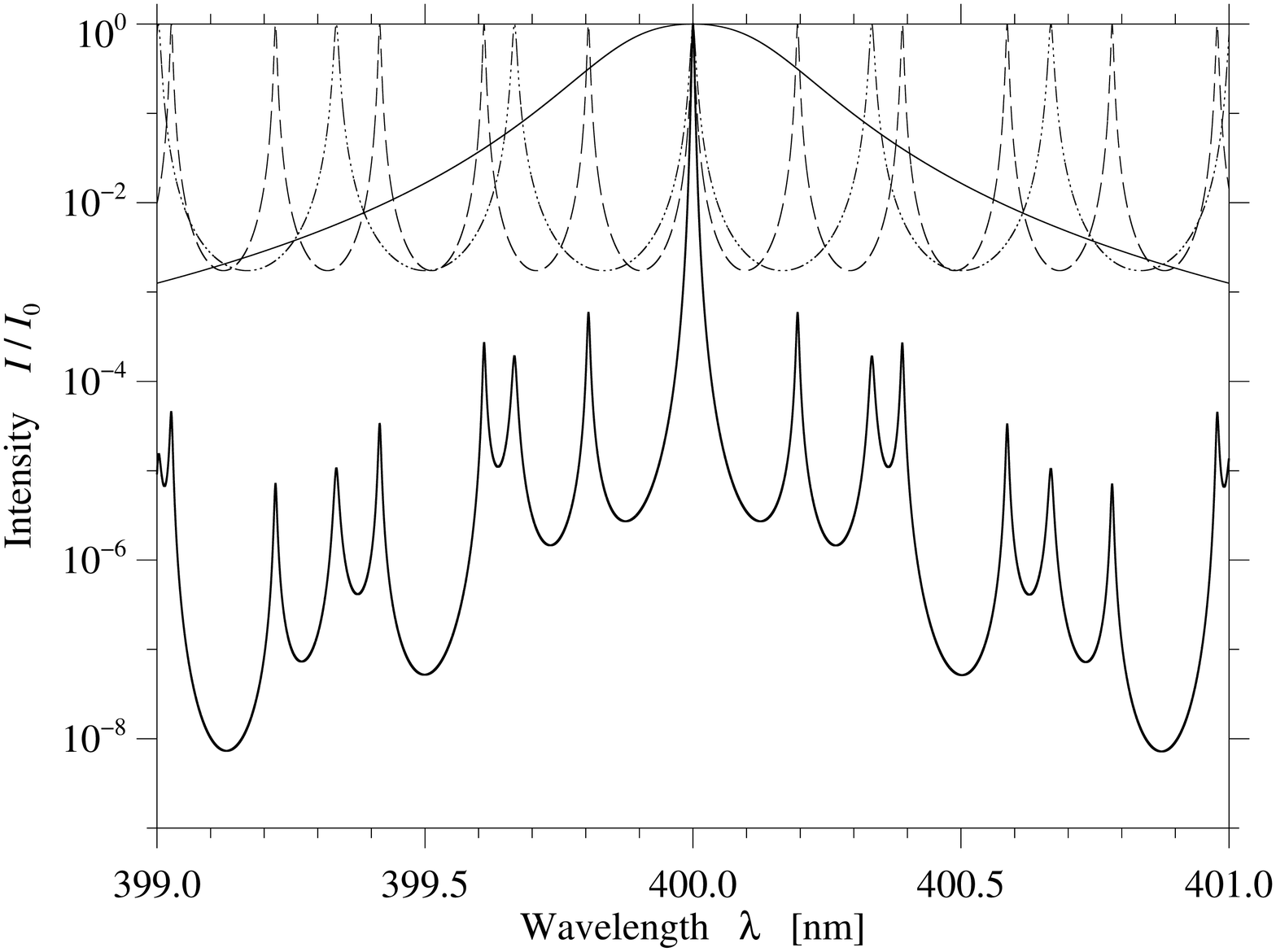}
\caption{Comparison of the GFPI (\textit{left})/BLISS (\textit{right})
    transmission profiles. The curves correspond to a narrow-band interference
    filter ($\mathrm{FWHM} = 0.3$~nm, \textit{thin solid line}), etalon~1 ($R =
    0.95/0.92$, $d = 1.4/0.41$~mm, $\mathrm{FWHM} = 2.1/5.2$~pm, \textit{thin
    dashed line}), etalon~2 ($R = 0.95/0.92$, $d = 1.1/0.24$~mm, $\mathrm{FWHM}
    = 2.7/8.9$~pm, \textit{thin dash-dotted line}), and all transmission curves
    combined ($\mathrm{FWHM} = 1.5/4.2$~pm, parasitic-light fraction 0.8/0.4\%,
    \textit{thick solid line}). The transmission profiles are normalized to
    unity at the central wavelength $\lambda_0 = 600.0/400.0$~nm.}
\label{FIG09}
\end{figure}

\subsection{Etalons}

Airy functions are quasi-periodic, i.e., in a dual etalon system their
characteristics parameters (reflectivity $R$ and plate separation $d$) have to
be carefully chosen to minimize parasitic light from neighboring
orders. Even if an optimal choice has been identified, narrow-band interference
filters (0.3--0.8~nm) have to be used to minimize the contributions by the
parasitic light. In general, an optimization of the characteristic FPI parameters has to
include realistic transmission curves for the interference filters. In
Fig.~\ref{FIG09}, we present the results of such a parameter study for BLISS and
compare them with the case of the GFPI. The characteristic parameters of the
BLISS etalons were chosen such that the parasitic light is minimized. The reflectivity
of etalons for imaging in the blue spectral region has to be lower to increase
the FWHM and to accommodate the smaller number of available photons. Therefore,
the rejection of parasitic light further away from the central wavelength
$\lambda_0$ becomes more important and narrower interference filters are
required. The combination of etalons with $d = 0.41$ and 0.24~mm ($\mathrm{FWHM}
= 5.2$ and 8.9~pm), will further increase the light level and result in a
theoretical spectral resolution of ${\cal R} \sim 100,000$.

%
%

\section{Data pipeline for imaging spectropolarimetry}

Analyzing data from imaging spectropolarimetry requires intimate knowledge about
seeing, telescope, Fabry-P\'erot etalons, detector systems, instrumental
polarization, and image restoration techniques. This knowledge about the entire
image formation process has to be encapsulated in a reliable and robust data
pipeline, which provides the user with well calibrated, self-describing data
suitable for further analysis using, for example, spectral line inversion
techniques.

The GFPI builds upon a two-decade heritage of Fabry-P\'erot interferometers
operated at the VTT. During this period the data reduction code bifurcated
significantly necessitating rewriting the code from scratch. The best features
of the individual codes were kept and in addition, the most time-consuming
algorithm were tuned for performance.

\subsection{Data processing and image restoration}

Imaging spectropolarimetry\cite{2011arXiv1111.5509P, 2011A&A...533A..21P} offers
the possibility to improve the recorded data beyond AO real-time correction.
Therefore, several state-of-the-art image restoration techniques will be part of
the GFPI data pipeline. At present, two of the most successful solar image
restoration tools are included, namely, the G\"ottingen speckle imaging and
deconvolution code\cite{1993PhDT.......155D, 2006A&A...454.1011P,
2007SoPh..241..411D} and Multi-Frame Blind
Deconvolution\cite{2002SPIE.4792..146L} (MFBD) as well as its extension to
multiple objects\cite{2005SoPh..228..191V} (MOMFBD). As a third option the
Kiepenheuer-Institute Speckle Interferometry Package\cite{2008A&A...488..375W,
2008SPIE.7019E..46W} (KISIP) will be included in the near future.

The data pipeline was developed in particular for data of the GFPI and BLISS.
However, its modular approach to image processing facilitates the integration of
other imaging spectropolarimeters. The I/O interface can be easily adapted.
Since image restoration is an integral part of the data pipeline, also images
from high-cadence, large-format CCD or CMOS cameras can be easily reduced.  The
information  about observatories, telescopes, instruments, and detectors  is
kept in configuration files, which can simply be modified for site-specific
needs.

The data pipeline is based on the Interactive Data Language (IDL). We adhere to
a slightly modified version of the IDL coding standard\cite{Excelis2006} to
provide the same ``touch-and-feel'' for all programs. This is enforced by using
the same variable names in all programs, where descriptive variable names are
the guiding principle rather than short names without meaning. The IDL source
code of the GFPI data pipeline is available to all GFPI users. However, the
hardware requirements go beyond that provided by typical work stations.
Nevertheless, scaled-down data processing is still possible on a desk-top
computer.

\subsection{Data formats and data conversion}

GFPI data are written in a native DaVis format, which uses on-the-fly image
compression to achieve high data rates while writing data to a RAID~0 harddisk
array. Narrow- and broad-band images are saved together in one file for each
simultaneous exposure. Imaging spectropolarimetric sequences are accompanied by
``set'' files, i.e., an ASCII file with all information describing the data and
observing modes. These compressed data are archived at AIP, whereas the
principle investigator (PI) of the observing run receives a copy of the data in
the Flexible Images Transport System\cite{Hanisch2001} (FITS) format using image
extensions\cite{Ponz1994}. Data from the GFPI data acquisition computer can be
transferred by FTP to servers for data storage and processing, which are located
at the VTT. The transfer lasts about the same time as taking the data. The raw
data are then converted to FITS. Both data types are stored finally to
LTO-Ultrium tapes with a storage capacity of about 800~GB.

\subsection{Quick-look data, on-line repositories, and automatic logbooks}

The data conversion step also includes several other tasks: (1) Image statistics
are computed and included in the headers of the image extensions. These values
are used in an heuristic error analysis to identify potential problems in the
data acquisition process early on. Together with  quantities describing the GFPI
performance, they are stored in a database to monitor the long-term stability of
the instrument and the quality of its data products. (2) Quick-look data
products can be computed in time spans that are comparable to the data
acquisition. Some functionality is already provided by the DaVis software
(visualization of spectral scans, contrast enhancement, or region-of-interest
manipulations). However, line-of-sight velocity maps or vector magnetograms have
to be computed off-line. Limiting the image restoration to just a simple
destretching of narrow- and broad-band images and taking averages at successive
wavelength positions already provides spatially resolved information about the
physical properties of the observed solar features. (3) Quick-look data are
offered in two formats, either as web pages or as logbooks in the Portable
Document Format (PDF), which both document each day's observing run. The web
pages with quick-look data will become public immediately, whereas the raw and
FITS data will be embargoed for a certain time (typically one year), if the data
are taken in the PI-mode.

\subsection{Observing modes}

Data from imaging spectropolarimetry can be analyzed in many different ways, for
example, different image restoration methods are available, noise in both
narrow- and broad-band images can be treated differently, or different schemes
for the polarimetric correction of the data can be applied. All parameters
describing a particular data processing scheme are saved in ``project'' files,
which also serve as templates for observing runs with similar characteristics.
This ensures that data products from different observing runs are comparable and
enables their use in databases, where a user expects to have data products of
the same quality.

Post-focus instruments at ground-based solar observatories are often operated in
PI-mode, where the PI of an observing run specifies the instrument setup and
observing sequences. The data are considered property of the PI and often only a
small fraction of data finds its way into scientific publications. Changing this
type of approach requires a concerted effort, both on the instrument-builder
side and on the part of the developers of data pipelines. The GFPI control
software offers a user interface for all standard data acquisition modes (dark,
flat-field, target, and pinhole frames as well as spectral line scans).
Interaction with the telescope, its subsystems, and other peripheral devices is
handled automatically for each task. The ``set'' files for each mode contain all
necessary information for further processing the data. The intermediate step of
converting the data to FITS format allows the users to use their own data
processing routines in addition to the GFPI data pipeline. A detailed account of
the data processing is beyond the scope of this conference proceedings and will
be deferred to a data release paper in a peer-reviewed journal.

%
%

\acknowledgements The 1.5-meter GREGOR solar telescope was built by a German
consortium under the leadership of the Kiepenheuer-Institut f{\"u}r Sonnenphysik
in Freiburg with the Leibniz-Institut f{\"u}r Astrophysik Potsdam and the
Max-Planck-Institut f{\"u}r Sonnensystemforschung in Katlenburg-Lindau as
partners, and with contributions by the Instituto de Astrof{\'{\i}}sica de
Canarias, the Institut f{\"u}r Astrophysik G{\"o}ttingen, and the Astronomical
Institute of the Academy of Sciences of the Czech Republic. CD was supported by
grant DE~787/3-1 of the Deutsche Forschungsgemeinschaft (DFG).

%
%


\begin{thebibliography}{10}

\bibitem{2003SPIE.4853..341S}
  {Scharmer}, G.~B., {Bjelksjo}, K., {Korhonen}, T.~K., {Lindberg}, B., and
  {Petterson}, B., ``{The 1-meter Swedish solar telescope},'' in [{\em
  Innovative telescopes and instrumentation for solar
  astrophysics}{\nolinebreak\hspace{0.1em}]},  {Keil}, S.~L. and {Avakyan},
  S.~V., eds., {\em Proc. SPIE} {\bf 4853},  341--350 (2003).

\bibitem{2007SoPh..243....3K}
  {Kosugi}, T., {Matsuzaki}, K., {Sakao}, T., {Shimizu}, T., {Sone}, Y.,
  {Tachikawa}, S., et al., ``{The Hinode (Solar-B) mission: an overview},'' {\em
  Sol. Phys.}~{\bf 243},  3--17 (2007).

\bibitem{2010ApJ...723L.127S}
  {Solanki}, S.~K., {Barthol}, P., {Danilovic}, S., {Feller}, A., {Gandorfer},
  A., {Hirzberger}, J., et al., ``{SUNRISE: instrument,
  mission, data, and first results},'' {\em ApJL}~{\bf 723},  L127--L133 (2010).

\bibitem{2007msfa.conf...39V}
  {Volkmer}, R., {von der L{\"u}he}, O., {Kneer}, F., {Staude}, J., {Balthasar},
  H., {Berkefeld}, T., et al., ``{New high resolution solar telescope GREGOR},''
  in [{\em Modern solar facilities -- advanced solar
  science}{\nolinebreak\hspace{0.1em}]},  {Kneer, F., Puschmann, K. G. \&
  Wittmann, A. D.}, ed.,  39 (2007).

\bibitem{2010SPIE.7733E..18V}
  {Volkmer}, R., {von der L{\"u}he}, O., {Denker}, C., {Solanki}, S.,
  {Balthasar}, H., {Berkefeld}, T., et al., ``{GREGOR telescope: start of
  commissioning},'' in [{\em Ground-based and airborne telescopes
  III}{\nolinebreak\hspace{0.1em}]},  {Stepp}, L.~M., {Gilmozzi}, R., and
  {Hall}, H.~J., eds., {\em Proc. SPIE} {\bf 7733},  77330K (2010).

\bibitem{2010AN....331..624V}
  {Volkmer}, R., {von der L{\"u}he}, O., {Denker}, C., {Solanki}, S.~K.,
  {Balthasar}, H., {Berkefeld}, T., et al., ``{GREGOR solar
  telescope: design and status},'' {\em Astronomische Nachrichten}~{\bf 331},
  624 (2010).

\bibitem{2012arXiv1202.4289S}
  {Schmidt}, W., {von der L{\"u}he}, O., {Volkmer}, R., {Denker}, C., {Solanki},
  S., {Balthasar}, H., et al., ``{The GREGOR solar
  telescope on Tenerife},'' ASP Conf. Ser., in press, {\em ArXiv e-prints, 2012arXiv1202.4289S}  (2012).

\bibitem{2006SPIE.6267E..10D}
  {Denker}, C., {Goode}, P.~R., {Ren}, D., {Saadeghvaziri}, M.~A., {Verdoni},
  A.~P., {Wang}, H., et al.,
  ``{Progress on the 1.6-meter New Solar Telescope at Big Bear Solar
  Observatory},'' in [{\em Ground-based and airborne
  telescopes}{\nolinebreak\hspace{0.1em}]},  {Stepp}, L.~M., ed., {\em Proc.
  SPIE} {\bf 6267},  62670A (2006).

\bibitem{2010SPIE.7733E..93C}
  {Cao}, W., {Gorceix}, N., {Coulter}, R., {Coulter}, A., and {Goode}, P.~R.,
  ``{First light of the 1.6 meter off-axis New Solar Telescope at Big Bear
  Solar Observatory},'' in [{\em Ground-based and airborne telescopes
  III}{\nolinebreak\hspace{0.1em}]},  {\em Proc. SPIE} {\bf 7733}, 773330-773330-8 (2010).






\bibitem{2008SPIE.7012E..16W}
  {Wagner}, J., {Rimmele}, T.~R., {Keil}, S., {Hubbard}, R., {Hansen}, E.,
  {Phelps}, L., et al., ``{Advanced Technology Solar Telescope: a
  progress report},'' in [{\em Ground-based and airborne telescopes
  II}{\nolinebreak\hspace{0.1em}]},  {Stepp}, L.~M. and {Gilmozzi}, R., eds.,
  {\em Proc. SPIE} {\bf 7012},  70120I (2008).


\bibitem{2010SPIE.7733E..15C}
  {Collados}, M., {Bettonvil}, F., {Cavaller}, L., {Ermolli}, I., {Gelly}, B.,
  {Grivel-Gelly}, C., et al., ``{European Solar Telescope: project status},'' in
  [{\em Ground-based and airborne telescopes III}{\nolinebreak\hspace{0.1em}]},
  {Stepp}, L.~M., R., G., and {Hall}, H.~J., eds., {\em Proc. SPIE} {\bf 7733},
   77330H (2010).

\bibitem{2011A&A...533A..21P}
  {Puschmann}, K.~G. and {Beck}, C., ``{Application of speckle and
  (multi-object) multi-frame blind deconvolution techniques on imaging and
  imaging spectropolarimetric data},'' {\em A\&A}~{\bf 533},  A21 (2011).

\bibitem{2011SoPh..268...57M}
  {Mart{\'{\i}}nez Pillet}, V., {Del Toro Iniesta}, J.~C.,
  {{\'A}lvarez-Herrero}, A., {Domingo}, V., {Bonet}, J.~A., {Gonz{\'a}lez
  Fern{\'a}ndez}, L., et al., ``{The
  Imaging Magnetograph eXperiment (IMaX) for the Sunrise balloon-borne solar
  observatory},'' {\em Sol. Phys.}~{\bf 268},  57--102 (2011).

\bibitem{2010SPIE.7733E..14R}
{Rimmele}, T.~R., {Wagner}, J., {Keil}, S., {Elmore}, D., {Hubbard}, R.,
  {Hansen}, E., et al., ``{The Advanced Technology Solar Telescope: beginning construction 
of the world's largest solar telescope},'' in [{\em Ground-based and
Aiborne Telescopes III}{\nolinebreak\hspace{0.1em}]},
{Stepp}, L.~M. and {Gilmozzi}, R., and {Helen}, H., eds., {\em Proc. SPIE} {\bf 7733},  77330G-17 (2010).

\bibitem{1992A&A...257..817B}
 {Bendlin}, C., {Volkmer}, R., and {Kneer}, F., ``{A new instrument for high
  resolution, two-dimensional solar spectroscopy},'' {\em A\&A}~{\bf 257},
  817--823 (1992).

\bibitem{1995A&A...304L...1V}
  {Volkmer}, R., {Kneer}, F., and {Bendlin}, C., ``{Short-period waves in
  small-scale magnetic flux tubes on the Sun.},'' {\em A\&A}~{\bf 304},  L1
  (1995).

\bibitem{2006A&A...451.1151P}
  {Puschmann}, K.~G., {Kneer}, F., {Seelemann}, T., and {Wittmann}, A.~D.,
  ``{The new G{\"o}ttingen Fabry-P{\'e}rot spectrometer for two-dimensional
  observations of the Sun},'' {\em A\&A}~{\bf 451}, 1151--1158 (2006).

\bibitem{2007msfa.conf...45P}
   {Puschmann}, K.~G., {Kneer}, F., {Nicklas}, H., and {Wittmann}, A.~D.,
  ``{From the ''G{\"o}ttingen'' Fabry-P\'erot interferometer to the GREGOR
  FPI},'' in [{\em Modern solar facilities -- advanced solar
  science}{\nolinebreak\hspace{0.1em}]},  {Kneer}, F., {Puschmann}, K.~G., and
  {Wittmann}, A.~D., eds., 45 (2007).

\bibitem{2008A&A...480..265B}
  {Bello Gonz{\'a}lez}, N. and {Kneer}, F., ``{Narrow-band full Stokes
  polarimetry of small structures on the Sun with speckle methods},'' {\em
  A\&A}~{\bf 480},  265--275 (2008).

\bibitem{2009IAUS..259..665B}
  {Balthasar}, H., {Bello Gonz{\'a}lez}, N., {Collados}, M., {Denker}, C.,
  {Hofmann}, A., {Kneer}, F., and {Puschmann}, K.~G., ``{A full-Stokes
  polarimeter for the GREGOR Fabry-P\'erot interferometer},'' in [{\em Cosmic
  Magnetic Fields: From Planets, to Stars and
  Galaxies}{\nolinebreak\hspace{0.1em}]},  {Strassmeier}, K.~G., {Kosovichev},
  A.~G., and {Beckman}, J.~E., eds., {\em IAU Symp.} {\bf 259},  665--666 (2009).

\bibitem{2011ASPC..437..351B}
  {Balthasar}, H., {Bello Gonz{\'a}lez}, N., {Collados}, M., {Denker}, C.,
  {Feller}, A., {Hofmann}, A., et al., ``{Polarimetry with GREGOR},'' in [{\em
  Solar polarization 6}{\nolinebreak\hspace{0.1em}]},  {Kuhn}, J.~R.,
  {Harrington}, D.~M., {Lin}, H., {Berdyugina}, S.~V., {Trujillo-Bueno}, J.,
  {Keil}, S.~L., and {Rimmele}, T., eds., {\em ASP Conf. Ser.} {\bf 437},  351
  (2011).

\bibitem{2010SPIE.7735E.217D}
  {Denker}, C., {Balthasar}, H., {Hofmann}, A., {Bello Gonz{\'a}lez}, N., and
  {Volkmer}, R., ``{The GREGOR Fabry-P\'erot interferometer: a new instrument
  for high-resolution solar observations},'' in [{\em Ground-based and airborne
  instrumentation for astronomy III}{\nolinebreak\hspace{0.1em}]},  {McLean},
  I.~S., {Ramsay}, S.~K., and {Takami}, H., eds., {\em Proc. SPIE} {\bf 7735},
  77356M--77356M--12 (2010).

\bibitem{2011arXiv1111.5509P}
  {Puschmann}, K.~G., {Balthasar}, H., {Bauer}, S.~M., {Hahn}, T., {Popow}, E.,
  {Seelemann}, T., et al., ``{The GREGOR
  Fabry-P\'erot Interferometer -- a new instrument for high-resolution
  spectropolarimetric solar observations},'' ASP Conf. Ser., in press, {\em ArXiv e-prints, 2011arXiv1111.5509P}  (2011).

\bibitem{2007msfa.conf...51S}
  {Strassmeier}, K.~G., {Woche}, M., {Granzer}, T., {Andersen}, M.~I.,
  {Schmidt}, W., and {Koubsky}, P., ``{Gregor@Night},'' in [{\em Modern solar
  facilities -- advanced solar science}{\nolinebreak\hspace{0.1em}]},  {Kneer,
  F., Puschmann, K. G. \& Wittmann, A. D.}, ed.,  51 (2007).

\bibitem{2008SPIE.7018E..52B}
  {Bettonvil}, F. C.~M., {Hammerschlag}, R.~H., {J{\"a}gers}, A. P.~L., and
  {Sliepen}, G., ``{Large fully retractable telescope enclosures still closable
  in strong wind},'' in [{\em Advanced Optical and Mechanical Technologies in
  Telescopes and Instrumentation}{\nolinebreak\hspace{0.1em}]},  {\em Proc.
  SPIE} {\bf 7018}, 70181N-70181N-9 (2008).

\bibitem{2010ApOpt..49G.155B}
  {Berkefeld}, T., {Soltau}, D., {Schmidt}, D., and {von der L{\"u}he}, O.,
  ``{Adaptive optics development at the German solar telescopes},'' {\em Appl.
  Optics}~{\bf 49},  G155 (2010).

\bibitem{2009CEAB...33..317H}
  {Hofmann}, A., {Rendtel}, J., and {Arlt}, K., ``{Toward polarimetry with
  GREGOR -- testing the GREGOR Polarimetric Unit},'' {\em Centr. Eur.
  Astrophys. Bull.}~{\bf 33},  317--325 (2009).

\bibitem{2008SPIE.7014E.198C}
  {Collados}, M., {Calcines}, A., {D{\'{\i}}az}, J.~J., {Hern{\'a}ndez}, E.,
  {L{\'o}pez}, R., and {P{\'a}ez}, E., ``{A high-resolution spectrograph for
  the solar telescope GREGOR},'' in [{\em Ground-based and airborne
  instrumentation for astronomy II}{\nolinebreak\hspace{0.1em}]},  {McLean},
  I.~S. and {Casali}, M.~M., eds., {\em Proc. SPIE} {\bf 7014},  70145Z (2008).


\bibitem{2007msfa.conf...55B}
{Beck}, C., {Mikurda}, K., {Bellot Rubio}, L.~R., {Kentischer}, T., and
  {Collados}, M., ``{Multi-wavelength observations at the German VTT on
  Tenerife},'' in [{\em Modern solar facilities - advanced solar
  science}{\nolinebreak\hspace{0.1em}]},  {Kneer}, F., {Puschmann}, K.~G., and
  {Wittmann}, A.~D., eds.,  55 (2007).

\bibitem{2011A&A...535A.129B}
{Beck}, C., {Rezaei}, R., and {Fabbian}, D., ``{Stray-light contamination and
  spatial deconvolution of slit-spectrograph observations},'' {\em A\&A}~{\bf
  535},  A129 (2011).

\bibitem{2012A&A...537A..80L}
{L{\"o}fdahl}, M.~G. and {Scharmer}, G.~B., ``{Sources of straylight in the
  post-focus imaging instrumentation of the Swedish 1-m Solar Telescope},''
  {\em A\&A}~{\bf 537},  A80 (2012).

\bibitem{2004A&A...423.1109A}
{Allende Prieto}, C., {Asplund}, M., and {Fabiani Bendicho}, P.,
  ``{Center-to-limb variation of solar line profiles as a test of NLTE line
  formation calculations},'' {\em A\&A}~{\bf 423},  1109--1117 (2004).

\bibitem{2007A&A...475.1067C}
{Cabrera Solana}, D., {Bellot Rubio}, L.~R., {Beck}, C., and {Del Toro
  Iniesta}, J.~C., ``{Temporal evolution of the Evershed flow in sunspots. I.
  Observational characterization of Evershed clouds},'' {\em A\&A}~{\bf 475},
  1067--1079 (2007).

\bibitem{2005PASP..117.1435D}
{Denker}, C. and {Tritschler}, A., ``{Measuring and Maintaining the Plate
  Parallelism of Fabry-P{\'e}rot Etalons},'' {\em PASP}~{\bf 117},  1435--1444
  (2005).

\bibitem{1993ApJ...414..345L}
{Lites}, B.~W., {Rutten}, R.~J., and {Kalkofen}, W., ``{Dynamics of the solar
  chromosphere. I - Long-period network oscillations},'' {\em ApJ}~{\bf 414},
  345--356 (1993).

\bibitem{2002A&A...389.1020R}
{Rouppe van der Voort}, L.~H.~M., ``{Penumbral structure and kinematics from
  high-spatial-resolution observations of Ca II K},'' {\em A\&A}~{\bf 389},
  1020--1038 (2002).

\bibitem{2005A&A...437.1159B}
{Beck}, C., {Schmidt}, W., {Kentischer}, T., and {Elmore}, D., ``{Polarimetric
  Littrow Spectrograph - instrument calibration and first measurements},'' {\em
  A\&A}~{\bf 437},  1159--1167 (2005).

\bibitem{2008A&A...479..213B}
{Beck}, C., {Schmidt}, W., {Rezaei}, R., and {Rammacher}, W., ``{The signature
  of chromospheric heating in Ca II H spectra},'' {\em A\&A}~{\bf 479},
  213--227 (2008).

\bibitem{1970SoPh...12...66S}
  {Sistla}, G. and {Harvey}, J.~W., ``{Fraunhofer lines without Zeeman
  splitting},'' {\em Sol. Phys.}~{\bf 12},  66--68 (1970).

\bibitem{1973SoPh...28....9H}
  {Harvey}, J.~W., ``{Fraunhofer lines with large Zeeman splitting},'' {\em Sol.
  Phys.}~{\bf 28},  9--13 (1973).

\bibitem{1993PhDT.......155D}
  {de Boer}, C.~R., {\em {Speckle-Interferometrie und ihre Anwendungen auf die
  Sonnenbeobachtung}}, PhD thesis, Universit{\"a}ts-Sternwarte G{\"o}ttingen,
  Germany (1993).

\bibitem{2006A&A...454.1011P}
  {Puschmann}, K.~G. and {Sailer}, M., ``{Speckle reconstruction of photometric
  data observed with adaptive optics},'' {\em A\&A}~{\bf 454},  1011--1019
  (2006).

\bibitem{2007SoPh..241..411D}
  {Denker}, C., {Deng}, N., {Rimmele}, T.~R., {Tritschler}, A., and {Verdoni},
  A., ``{Field-dependent adaptive optics correction derived with the spectral
  ratio technique},'' {\em Sol. Phys.}~{\bf 241},  411--426 (2007).

\bibitem{2002SPIE.4792..146L}
  {L{\"o}fdahl}, M.~G., ``{Multi-frame blind deconvolution with linear equality
  constraints},'' in [{\em Image reconstruction from incomplete data
  II}{\nolinebreak\hspace{0.1em}]},  {Bones}, P.~J., {Fiddy}, M.~A., and
  {Millane}, R.~P., eds., {\em Proc. SPIE} {\bf 4792},  146--155 (2002).

\bibitem{2005SoPh..228..191V}
  {van Noort}, M., {Rouppe van der Voort}, L., and {L{\"o}fdahl}, M.~G.,
  ``{Solar image restoration by use of multi-frame blind de-convolution with
  multiple objects and phase diversity},'' {\em Sol. Phys.}~{\bf 228},  191--215
  (2005).

\bibitem{2008A&A...488..375W}
  {W{\"o}ger}, F., {von der L{\"u}he}, O., and {Reardon}, K., ``{Speckle
  interferometry with adaptive optics corrected solar data},'' {\em A\&A}~{\bf
  488},  375--381 (2008).

\bibitem{2008SPIE.7019E..46W}
  {W{\"o}ger}, F. and {von der L{\"u}he}, O., ``{KISIP: a software package for
  speckle interferometry of adaptive optics corrected solar data},'' in [{\em
  Advanced software and control for astronomy II}{\nolinebreak\hspace{0.1em}]},
   {Bridger}, A. and {Radziwill}, N.~M., eds., {\em Proc. SPIE} {\bf 7019},
  70191E (2008).

\bibitem{Excelis2006}
{Excelis Visual Information Solutions}, ``{One proposal for an IDL coding
  standard},'' Article ID 4120 (2006).

\bibitem{Hanisch2001}
  {Hanisch}, R.~J., {Farris}, A., {Greisen}, E.~W., {Pence}, W.~D.,
  {Schlesinger}, B.~M., {Teuben}, P.~J., et al.,
  ``{Definition of the Flexible Image Transport System (FITS)},'' {\em
  A\&A}~{\bf 376},  359--380 (2001).

\bibitem{Ponz1994}
  {Ponz}, J.~D., {Thompson}, R.~W., and {Munoz}, J.~R., ``{The FITS image
  extension},'' {\em A\&ASS}~{\bf 105},  53--55 (1994).

\end{thebibliography}

\end{document}